\documentclass[twocolumn,traditabstract, longauth]{aa}
\DeclareUnicodeCharacter{02BC}{'}
\usepackage{graphicx}
\usepackage[hyperindex,breaklinks=true, colorlinks, citecolor=blue]{hyperref} 
\usepackage{tablefootnote}
\usepackage{txfonts}
\usepackage{xcolor}
\usepackage{multirow}
\usepackage{ulem}
\usepackage{rotating}
\usepackage{orcidlink}
\usepackage{lscape}
\usepackage{booktabs}

\begin{document}
\newcommand{\dco}{DCO$^+$}
\newcommand{\ntwod}{N$_2$D$^+$}
\newcommand{\source}{AG286.0716$-$1.8229}


\title{ALMAGAL VIII. Early phases of triggered star formation in source AG286.0716$-$1.8229}
\author{C.~Mininni\inst{\ref{rome}}\orcidlink{0000-0002-2974-4703}\and 
S.~Molinari\inst{\ref{rome}}\orcidlink{0000-0002-9826-7525} \and
W.~J.~Kim\inst{\ref{unikoeln},\ref{bonn}}\and
E.~Schisano\inst{\ref{rome}}\orcidlink{0000-0003-1560-3958} \and
F.\ Fontani\inst{\ref{arcetri}, \ref{mpe}, \ref{lerma}} \and
A.~Traficante\inst{\ref{rome}}\orcidlink{0000-0003-1665-6402} \and
A.~Nucara\inst{\ref{rome}, \ref{torvergata}}\orcidlink{0009-0005-9192-5491} \and
A.~Coletta\inst{\ref{rome}}\orcidlink{0000-0001-8239-8304} \and
H.~T.~Lee\inst{\ref{unikoeln}}\and
\'A.~S\'anchez-Monge\inst{\ref{icecsic}, \ref{ieec}}\orcidlink{0000-0002-3078-9482} \and
M.~Benedettini\inst{\ref{rome}}\orcidlink{0000-0002-0560-3172} \and
D.~Elia\inst{\ref{rome}}\orcidlink{0000-0002-9120-5890} \and
S.~Pezzuto\inst{\ref{rome}}\orcidlink{0000-0001-7852-1971} \and
V.~M. Pelkonen\inst{\ref{rome}}\and
P.~Schilke\inst{\ref{unikoeln}}\orcidlink{0000-0002-1730-8832} \and
C.~Battersby\inst{\ref{connecticut}}\orcidlink{0000-0002-6073-9320} \and
P.~T.~P.~Ho\inst{\ref{asiaa}, \ref{hawaii}}\orcidlink{0000-0002-3412-4306} \and
M.~T.~Beltr\'an\inst{\ref{arcetri}}\orcidlink{0000-0003-3315-5626} \and
H.~Beuther\inst{\ref{mpia}}\orcidlink{0000-0002-1700-090X} \and
G.~A.~Fuller\inst{\ref{manchester}, \ref{unikoeln}}\orcidlink{0000-0001-8509-1818} \and
B.~Jones\inst{\ref{unikoeln}}\orcidlink{0000-0002-0675-0078} \and
R.~S.~Klessen\inst{\ref{ita}, \ref{uniheidelberg},\ref{cfa},\ref{radcliffeHav}}\orcidlink{0000-0002-0560-3172} \and
Q.~Zhang\inst{\ref{cfa}}\orcidlink{0000-0003-2384-6589} \and
S.~Walch\inst{\ref{unikoeln}, \ref{datacologne}} \and
Y.~Tang\inst{\ref{asiaa}} \and
A.~Ahmadi\inst{\ref{astron}}\orcidlink{0000-0003-4037-5248} \and
J. Allande\inst{\ref{arcetri},\ref{unifirenze}}\and
A.~Avison\inst{\ref{skaUK}, \ref{manchester}, \ref{almaUK}}\orcidlink{0000-0002-2562-8609} \and
J.~Ballesteros-Paredes\inst{\ref{morelia}}\and
L.~Bronfman\inst{\ref{unichile}}\orcidlink{0000-0002-9574-8454}\and
C.~L.~Brogan\inst{\ref{nraoVA}}\orcidlink{0000-0002-6558-7653} \and
F. De Angelis\inst{\ref{rome}}\orcidlink{0009-0002-6765-7413}\and 
E.~Rodrigues da Costa\inst{\ref{unikoeln}}\and
P. Hennebelle\inst{\ref{saclay}}\and
T.~R.~Hunter\inst{\ref{nraoVA}}\orcidlink{0000-0001-6492-0090} \and
K.~G.~Johnston\inst{\ref{lincoln}}\orcidlink{0000-0003-4509-1180} \and
K.~T.~Kim\inst{\ref{korea1},\ref{korea2}}\and
P.~Klaassen\inst{\ref{edimburgo}}\and
R.~Kuiper\inst{\ref{duisburg}}\orcidlink{0000-0003-2309-8963} \and
C.~-Y.~Law\inst{\ref{arcetri}}\and
D.~C.~Lis\inst{\ref{jplcaltech}}\orcidlink{0000-0002-0500-4700} \and
S. Liu\inst{\ref{rome}}\orcidlink{0000-0001-7680-2139}\and
L.~Moscadelli\inst{\ref{arcetri}}\orcidlink{0000-0002-8517-8881} \and
T.~M\"oller\inst{\ref{unikoeln}}\orcidlink{0000-0002-9277-8025} \and
K.~L.~J.~Rygl\inst{\ref{bologna}}\orcidlink{0000-0003-4146-9043} \and
P.~Sanhueza\inst{\ref{tokyo}}\orcidlink{0000-0002-7125-7685}\and
J.~D.~Soler\inst{\ref{vienna},\ref{rome}}\orcidlink{0000-0002-0294-4465} \and
Y.~N.~Su\inst{\ref{asiaa}}\and
L. Testi\inst{\ref{u-bo}}\and
F.~F.~S.~van der Tak\inst{\ref{sron}, \ref{kapteyn}}\orcidlink{0000-0002-8942-1594} \and
T.~Zhang\inst{\ref{zhejiang}, \ref{unikoeln}}\orcidlink{0000-0002-1466-3484}\and
H.~Zinnecker\inst{\ref{autchile}}\and
J.~Wallace\inst{\ref{connecticut}}
%
%
}

\institute{
\label{rome}INAF-Istituto di Astrofisica e Planetologia Spaziale, Via Fosso del Cavaliere 100, I-00133 Roma, Italy
\and
\label{unikoeln}I.\ Physikalisches Institut, Universit\"{a}t zu K\"{o}ln, Z\"{u}lpicher Str.\ 77, D-50937 K\"{o}ln, Germany
\and
\label{bonn}Max-Planck-Institut f\"{u}r Radioastronomie, Auf dem H\"{u}gel 69, 53121 Bonn, Germany
\and
\label{arcetri}INAF-Osservatorio Astrofisico di Arcetri, Largo E.\ Fermi 5, I-50125 Firenze, Italy
\and
Max-Planck-Institute for Extraterrestrial Physics (MPE), Garching bei M\"unchen, Germany \label{mpe} 
\and
Laboratory for the study of the Universe and eXtreme phenomena (LUX), Observatoire de Paris, Meudon, France\label{lerma} 
\and
\label{icecsic}Institut de Ci\`encies de l'Espai (ICE), CSIC, Campus UAB, Carrer de Can Magrans s/n, E-08193, Bellaterra (Barcelona), Spain\\ \email{asanchez@ice.csic.es}
\and
\label{ieec}Institut d'Estudis Espacials de Catalunya (IEEC), E-08860, Castelldefels (Barcelona), Spain
\and
\label{torvergata}Dipartimento di Fisica, Universit\`a di Roma Tor Vergata, Via della Ricerca Scientifica 1, I-00133 Roma, Italy
\and
\label{connecticut}University of Connecticut, Department of Physics, 2152 Hillside Road, Unit 3046 Storrs, CT 06269, USA
\and
\label{asiaa}Institute of Astronomy and Astrophysics, Academia Sinica, 11F of ASMAB, AS/NTU No.\ 1, Sec.\ 4, Roosevelt Road, Taipei 10617, Taiwan
\and
\label{hawaii}East Asian Observatory, 660 N.\ A'ohoku, Hilo, Hawaii, HI 96720, USA
\and
\label{mpia}Max Planck Institute for Astronomy, K\"onigstuhl 17, D-69117 Heidelberg, Germany
\and
\label{manchester}Jodrell Bank Centre for Astrophysics, Oxford Road, The University of Manchester, Manchester M13 9PL, UK
\and
\label{ita}Universit\"at Heidelberg, Zentrum f\"ur Astronomie, Institut f\"ur Theoretische Astrophysik, Heidelberg, Germany
\and
\label{uniheidelberg}Universit\"at Heidelberg, Interdisziplin\"ares Zentrum f\"ur Wissenschaftliches Rechnen, Heidelberg, Germany
\and
\label{cfa}Center for Astrophysics $|$ Harvard \& Smithsonian, 60 Garden St, Cambridge, MA 02138, USA
\and
\label{radcliffeHav}Elizabeth S. and Richard M. Cashin Fellow at the Radcliffe Institute for Advanced Studies at Harvard University, 10 Garden Street, Cambridge, MA 02138, U.S.A.
\and
\label{datacologne}Center for Data and Simulation Science, University of Cologne, Germany
\and
\label{astron}ASTRON, Netherlands Institute for Radio Astronomy, Oude Hoogeveensedijk 4, Dwingeloo, 7991 PD, The Netherlands
\and
\label{unifirenze}Dipartimento di Fisica e Astronomia, Universit\`a degli Studi di Firenze, Via G.\ Sansone 1,I-50019 Sesto Fiorentino, Firenze, Italy
\and
\label{skaUK}SKA Observatory, Jodrell Bank, Lower Withington, Macclesfield, SK11 9FT, UK
\and
\label{almaUK}UK ALMA Regional Centre Node, M13 9PL, UK
\and
\label{morelia}Universidad Nacional Aut\`onoma de M\`exico, Instituto de Radioastronom\`ia y Astrofísica, Apartado Postal 3-72, 58089 Morelia Michoac\`an, M\`exico
\and
\label{unichile}Departamento de Astronomía, Universidad de Chile, Casilla 36-D, Santiago, Chile
\and
\label{nraoVA}National Radio Astronomy Observatory, 520 Edgemont Road, Charlottesville VA 22903, USA
\and
\label{lincoln}School of Engineering and Physical Sciences, Isaac Newton Building, University of Lincoln, Brayford Pool, Lincoln, LN6 7TS, United Kingdom
\and
Universit\'e Paris-Saclay, Universit\'e Paris-Cit\'e, CEA, CNRS, AIM, 91191 Gif-sur-Yvette, France\label{saclay}
\and
\label{korea1}Korea Astronomy and Space Science Institute, 776 Daedeokdae-ro, Yuseong-gu, Daejeon 34055, Republic of Korea
\and
\label{korea2}University of Science and Technology, Korea (UST), 217 Gajeong-ro, Yuseong-gu, Daejeon 34113, Republic of Korea 
\and
\label{edimburgo}UK Astronomy Technology Centre, Royal Observatory Edinburgh, Blackford Hill, Edinburgh EH9 3HJ, UK
\and
\label{duisburg}Faculty of Physics, University of Duisburg-Essen, Lotharstra{\ss}e 1, D-47057 Duisburg, Germany
\and
\label{jplcaltech}Jet Propulsion Laboratory, California Institute of Technology, 4800 Oak Grove Drive, Pasadena, CA 91109, USA
\and
\label{bologna}INAF-Istituto di Radioastronomia \& Italian ALMA Regional Centre, Via P. Gobetti 101, I-40129 Bologna, Italy
\and
\label{tokyo}Department of Astronomy, School of Science, The University of Tokyo, 7-3-1 Hongo, Bunkyo, Tokyo 113-0033, Japan
\and
\label{vienna}Department of Astrophysics, University of Vienna, Turkenschanzstrasse 17, A-1180 Vienna, Austria
\and
\label{u-bo}Dipartimento di Fisica e Astronomia, Alma Mater Studiorum - Universit\`a di Bologna
\and
\label{sron}SRON Netherlands Institute for Space Research, Landleven 12, 9747 AD Groningen, The Netherlands
\and
\label{kapteyn}Kapteyn Astronomical Institute, University of Groningen, The Netherlands
\and
\label{zhejiang}Zhejiang Laboratory, Hangzhou 311100, P.R.China
\and
\label{autchile}Universidad Autonoma de Chile, Pedro de
Valdivia 425, 9120000 Santiago de Chile, Chile
}


   \date{Received ; accepted }



  \abstract
   {Several theoretical and observational studies have shown that new waves of triggered star-formation can be induced by the feedback from newly formed massive protostars, due to the expansion of \ion{H}{ii} regions. 
   }
   {We used the millimeter dust continuum data of the ALMAGAL survey and the Anderson et al. 2014 catalog of \ion{H}{ii} regions to search for signatures of possible triggered star formation at its onset, and selected one ALMAGAL source for ALMA follow-up observations. 
   }
   {For this study we selected the source \source, in which six cores were detected at a resolution of $\sim7600$\,au, but only two at a higher resolution. The 4 cores not detected at higher resolution are prestellar core candidates. We used archival data from the SARAO Meerkat Galactic Plane Survey and Rapid Askap Continuum Survey to confirm whether an \ion{H}{ii}   region is present in the field. We observed the source with with ALMA in Band 4, covering the emission of \dco (2$-$1), \ntwod (2$-$1), DCN (2$-$1), and CH$_3$CCH (9--8), to infer from the chemical composition and  temperature estimates whether these cores are in an early phase of the star-formation process, which allows us to classify them as prestellar core. The new Band 4 continuum image revealed additional three cores outside of the ALMAGAL field of view, for a total of 9 cores in the region, 8 of which are located along an arch of radius $\sim0.75$\,pc. We also used the continuum and molecular emission to estimate the mass of these cores, to determine whether they are gravitationally bound or transient objects.}
   {We have derived a spectral index between --0.14 and --0.4, in the frequency range of 0.8--1.6 GHz for the candidate\ion{H}{ii} region,  which is consistent with optically thin free-free emission. The \ion{H}{ii} region spatially coincides (as projected on the plane of the sky) with the position of one of the cores, which in the new ALMA Band 4 observations does not have a compact morphology. The \ntwod and \dco emission peaks at the position of three out of four prestellar core candidates, and the overall emission in both the line and the continuum, which extends outside of the FOV of the ALMAGAL observations, reveals a clear arched shape centered at the \ion{H}{ii} region. We were able to estimate the gas temperature of only the most evolved core, using the CH$_3$CCH emission. The resulting value of $39\pm5$\,K is also an upper limit for the other cores. Using plausible temperature ranges for each core, based on the information from chemical tracers and the dust continuum, we derived mass ranges for the cores ($\sim2-16\,$M$_{\odot}$) as well as ranges for the virial parameter ($\sim0.3-5$). All the cores along the arch have virial parameters consistent with bound objects, with only one exception. } 
   { Comparing the typical separation and mass of the cores  with those expected in the case of the collect and collapse scenario and with the thermal Jean length and mass, the best agreement is found with the characteristic scales in the case of triggered star formation. The same methodology can be applied to further ALMAGAL sources to identify optimal targets to study a larger sample of prestellar cores in an environment affected by the presence of an \ion{H}{ii} region.}

\keywords{ Astrochemistry -- ISM: molecules --Stars: formation -- ISM: \ion{H}{ii} regions }
\titlerunning{ALMAGAL }
\authorrunning{Mininni et al.}
    \maketitle
\section{Introduction} \label{sec:intro}
The formation of OB stars has a deep impact on the environment in which they are born, since they 
ionize their surrounding medium and create \ion{H}{ii} regions. The expansion of these \ion{H}{ii} regions could trigger a new wave of star-formation. During the expansion of the ionizing front, previously diffuse homogeneous material can be collected, compressed, and it then collapses due to gravity \citep[collect and collapse scenario]{elmegreen1977}. Alternatively, the expanding \ion{H}{ii} regions can compress preexisting inhomogeneous materials containing dense structures, making them reach the condition for their collapse \citep{bertoldi1989}, or a combination of the two effects could be at work \citep{walch2015}. \\ \indent 
The triggering effect of \ion{H}{ii} regions for new star-formation has been predicted by numerical simulations \citep[e.g.][]{hosokawa2005,dale2007}, and  confirmed in various observational cases \citep[e.g.][]{zavagno2006, zavagno2007, zavagno2010, deharveng2008,  fontani2012,Liu2015,palmeirim2017,zhou2020,sharma2022}. However, the majority of the observational studies analyzed evolved \ion{H}{ii} regions and looked at the distribution of already formed young stellar objects (YSOs) or the CO emission  \citep{elia2007VelaD,thompson2012,kendrew2012}, while only a few cases analyzed young star-forming material using higher density tracers \citep[e.g.][]{fontani2012, tackenberg2013, kendrew2016, palmeirim2017, Zhang2024ATOMS, kim2025}. The details of this triggered star formation, and the possible differences between star formation triggered by the presence of an \ion{H}{ii} regions and the scenario without feedback from already formed massive stellar objects, have not been fully explored, especially at the onset of the process. Our goal is to investigate the early stage of the process in the two scenario.  \\ \indent For this reason, we used the ALMA evolutionary study of high-mass protocluster formation in the GALaxy (ALMAGAL) survey \citep{ALMAGAL1} to search for sites of the early phase of triggered star-formation. ALMAGAL observed more than 1000 high-mass clumps in different environments in our Galaxy, and at different evolutionary stages with the Atacama Large Millimeter/Submillimeter Array (ALMA) in Band 6, down to a resolution of $\sim1000\,$au \citep{ALMAGAL2, ALMAGAL3}. To search for cores in the earliest evolutionary stage, we compared the continuum emission of the ALMAGAL field at the full resolution of $\sim1000\,$au\footnote{this is the typical value for the whole ALMAGAL sample. For source \source\, it is 1400\,au.} and at an intermediate resolution of $\sim5000\,$au\footnote{this is the typical value for the whole ALMAGAL sample. For source \source\, it is 7600\,au.} (see Section 2.1 and \citealt{ALMAGAL2} for more details). 
Comparing the two sets of images it is possible that a compact source in the maps at $\sim 5000\,$au is not identified above the noise limits in the $\sim1000\,$au if the radial profile of the source is flat, like in the case of prestellar cores \citep[e.g.][]{caselli2019flat}. In such a case, the same flux will be spread over a larger number of smaller beams in the full resolution image, and can be undetected. This will not happen for more evolved cores that will have a peaked radial profile structure.  Therefore, to search for candidate prestellar cores we looked for compact sources detected at $\sim5000\,$au, but not matched with any core at $\sim1000\,$au, which we refer to hereafter as disappearing cores. Moreover, to have good candidates for this pilot study we searched for \ion{H}{ii} regions or candidate \ion{H}{ii} regions in the ALMAGAL fields in the catalog of \cite{anderson2014} closer than 1\,pc to one or more disappearing cores. These disappearing cores are candidates prestellar cores in regions influenced by the presence of a possible \ion{H}{ii} region.
 \\ \indent Among the $\sim$1000 regions in ALMAGAL, we identified \source\, as a very promising region to study the presence of very early stage dense cores in the vicinity of an expanding \ion{H}{ii} region, which has four disappearing cores as will be shown in detail in Section 3. The aim of this study is to confirm the nature of the disappearing cores as bound objects in a pristine phase of the star-formation process, and investigate the scenario of triggered star-formation in the source using new observations in ALMA Band 4 of deuterated molecular species and continuum archival data at centimeter wavelength. In fact, in the early stages of star formation in regions with temperature below 20\,K and density above $\sim10^5\,$cm$^{-3}$ the  deuteration process takes place \citep[e.g.][]{caselli2012, fontanideuteration,fontani2014timescale}. This process leads to a significant enhancement of the abundance of several deuterated species. Some of the species affected are \dco and \ntwod, included in our observations.  Once this scenario is tested, the method could be expanded to other regions to improve the statistics on these objects and have a base to evaluate systemic differences between the star-formation process in the presence of \ion{H}{ii} regions and non-triggered star formation, from the time of their onset. \\
 In Section 2 we present the dataset of the observations; in Section 3 we analyze the continuum emission in different bands and the line emission from new ALMA Band 4 observations; in Section 4 we discuss the results; in Section 5 we discussed whether the formation of the cores analyzed is really triggered by the presence of an \ion{H}{ii} region; in Section 6 we summarize our main results.

\begin{figure*}
    \centering
    \includegraphics[width=18.0cm, trim={4.5cm, 8cm, 0cm, 0}, clip=True]{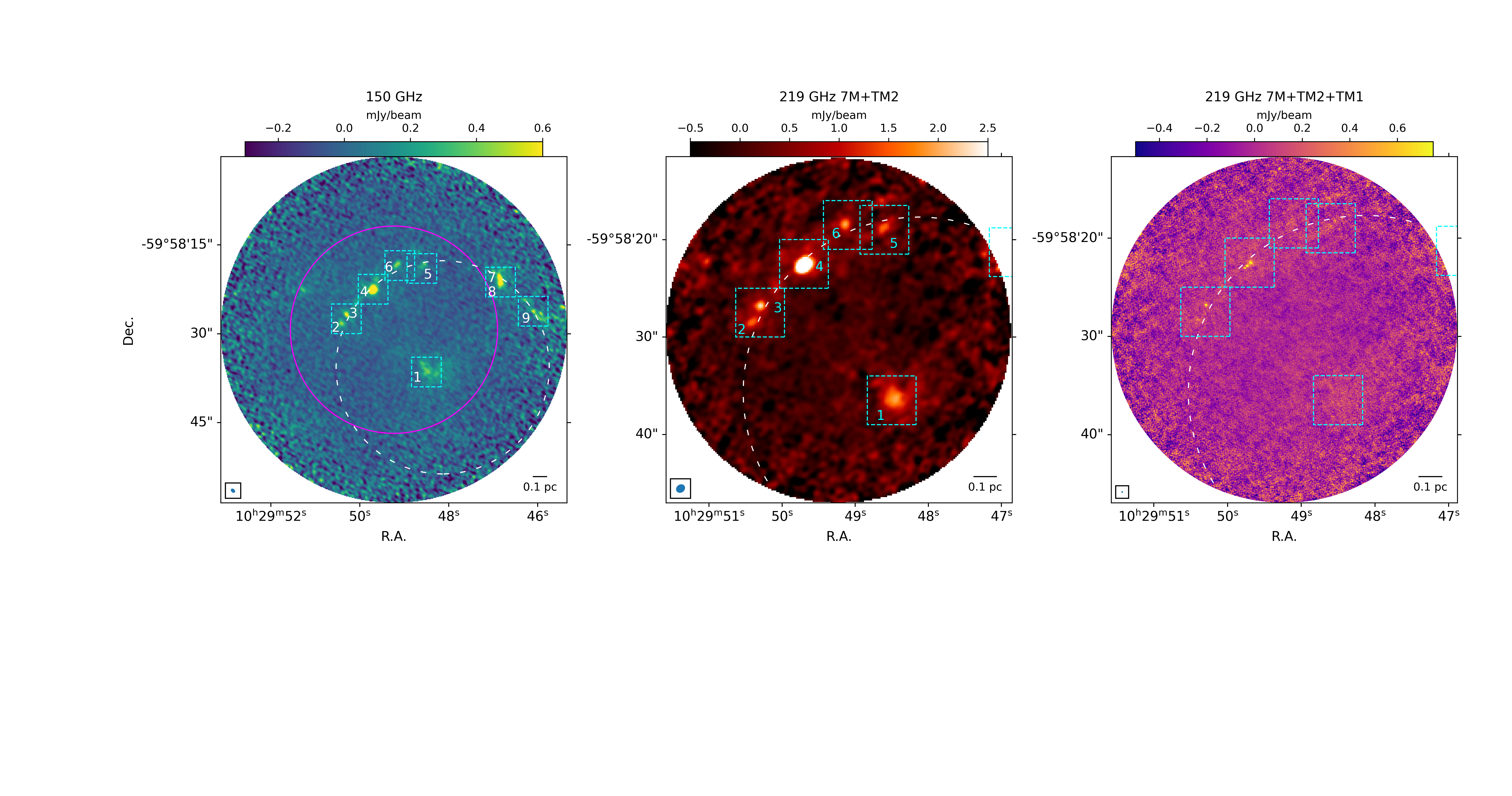}\\
     \includegraphics[width=18.0cm , trim={0cm, 0cm, 0cm, 1cm}, clip=True]{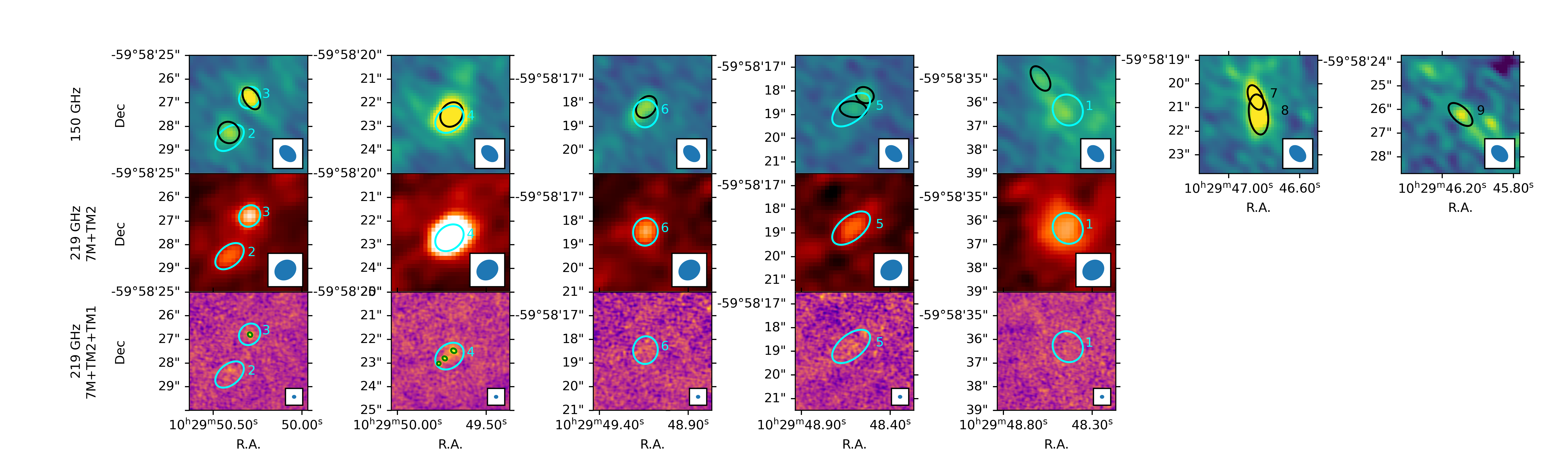}
    \caption{Top left panel: continuum emission at 150 GHz; the magenta circle represents the FOV of the ALMAGAL observations, while the dashed white circle visually represent the arch-shaped region where the continuum cores are detected; top middle panel: ALMAGAL \textsc{7m+tm2} continuum emission, the cyan boxes are the footprint of the zoom-ins in the lower panels; top right panel: ALMAGAL \textsc{7m+tm2+tm1} continuum emission. The white dashed circle in all the three top panels visually represent the 0.75\,pc radius circle along which all cores, except core \#1, are aligned. Lower panel: zoom-ins of the regions delimited in cyan in the top middle panel, where we kept the same colorscale for the three different datasets. The cyan ellipses represent the compact sources identified with CuTEx in the ALMAGAL \textsc{7m+tm2} data, with their identification number reported in cyan, while the black ellipses and the green ellipses are the compact sources identified with CuTEx in the ALMA Band 4 data and the ALMAGAL \textsc{7m+tm2+tm1} data, respectively.}
    \label{fig:3continuum}
\end{figure*}
\section{Observations}

\begin{table*}[]
    \centering
    \caption{Summary of the data analyzed in this work, presented in Sect. 2.}
    \begin{tabular}{lcccc}
    \hline\hline
        name & instrument & frequency & \multicolumn{2}{c}{beam}   \\
        & & \tiny{[GHz]} & \tiny{["]} & \tiny{[au]}\\
         \hline
        ALMA Band 4 & ALMA & 150& 0.70 & 6000\\
        ALMAGAL \textsc{7m+tm2} & ALMA & 219 & 0.89& 7600 \\
        ALMAGAL \textsc{7m+tm2+tm1} & ALMA & 219 & 0.17& 1400 \\
        SMGPS & Meerkat & 0.9-1.6 & 8 & 70000 \\
        RACS & ASKAP & 0.88 & 24 & 200000\\
         & & 1.35-1.65 & 10 & 86000\\
         \hline\hline
         
    \end{tabular}
    \label{tab:listdata}
    \label{tab:alldata}
\end{table*}
Table \ref{tab:listdata} gives a small summary of the dataset analyzed in Section 3 and described more in details in the following subsections.
\subsection{ALMA Band 4 Observations}
Our new observations were carried out with ALMA the 16th of August 2024 in configuration C-4 (Project 2023.1.01386.S, PI C. Mininni). The source \source\ has coordinates $l=286.0716^{\circ}$ and $b=-1.8229^{\circ}$, and an heliocentric distance of 8.5 kpc \citep{ALMAGAL1}. The source has been observed with elevation between $\sim45^{\circ}$ and $\sim53^{\circ}$ for a total of $\sim3.2\,$hrs on source. The sources selected for the phase and water vapor radiometer calibrations were J1032--5917 and J1019--6047.\\ \indent 
The continuum and cubes analyzed in this paper are the products of the standard pipeline and imaging (version 2023.1.0.124, \citealt{hunter2023}) provided by ALMA using CASA version 6.5.4.9 \citep{casa2022} using a robust value of 0.5. The linear resolution of the observations is of 0.83"$\times$0.59" ($\sim6000\,$au) and the maximum recoverable scale is 7.5" ($\sim0.3$\,pc), while the FOV is of 58.4".  We observed the source \source\ in four spectral windows in Band 4, centered around the emission lines of \dco (2$-$1) (144.077 GHz), \ntwod (2$-$1) (154.217 GHz), DCN (2$-$1) (144.828 GHz), and CH$_3$CCH 9-8 ($K=0$: 153.817 GHz). Each spectral window has a bandwidth of 117.19 MHz and a spectral resolution of 0.098 MHz ($\sim$0.2\,km\,s$^{-1}$). 
\subsection{ALMA Band 6 Observations}
The observations in ALMA Band 6 are part of the ALMAGAL large program (project
code: 2019.1.00195.L in Cycle 7; P.I. Molinari, Schilke, Battersby, Ho; \citealt{ALMAGAL1, ALMAGAL2}) that was observed between October 2019 and July 2022. The target name in ALMA archive (internal ID) for source \source\, is 646390. To achieve the requested linear resolution without losing information on diffuse material at the clump scale, each target was observed with the 7-m array (hereafter \textsc{7m}) and with two configurations of the main ALMA array.  \source\, was observed in configurations C-3 (\textsc{tm2}) and C-6 (\textsc{tm1}). 
The ALMAGAL team developed a dedicated pipeline to obtain the final cubes, using joint deconvolution of the various configurations observed, and that includes self-calibration 
for the most intense sources.  A detailed description of the data reduction, together with a quality assessment of the final products, is presented in \citet{ALMAGAL2}. The ALMAGAL pipeline produced continuum maps and cubes obtained by the joint deconvolution of \textit{\sc 7m} and \textit{\sc tm2} data only (hereafter \textit{\sc 7m+tm2}), which reach a linear resolution of $\sim7600\,$au ($\sim0.89$"), and of all the available configurations (hereafter  \textit{\sc 7m+tm2+tm1}) that reach the maximum resolution of $\sim1400\,$au ($\sim0.17$") for source \source. The resolution reached for the images and cubes of \source\, is worse than the typical average resolution in the ALMAGAL sample ($\sim5000$\,au and $\sim1000\,$au), since this is one of the farthest sources in the sample at 8.5\,kpc heliocentric distance. The diameter of the field of view (FOV) of the observations is 35.4" (down to the
30\% primary beam response).\\ \indent
The spectral setup of the observations covers four spectral windows in total: two spectral windows with 
a 1.9 GHz bandwidth and a spectral resolution of 0.98 MHz ($\sim1.3\,$km\,s$^{-1}$) centered at 217.8 GHz and 220.0 GHz, and two higher resolution spectral windows with a 0.47 GHz bandwidth and spectral resolution of 0.24 MHz ($\sim0.3\,$km\,s$^{-1}$) centered at 218.3 GHz and 220.6 GHz, respectively. The spectral frequency ranges observed include the frequency of emission of SiO $(5-4)$ transition, with rest frequency of 217.105\,GHz.

\begin{figure}
    \centering
    \includegraphics[width=8.5cm, trim={21.3cm, 0cm, 0cm, 0}, clip=True]{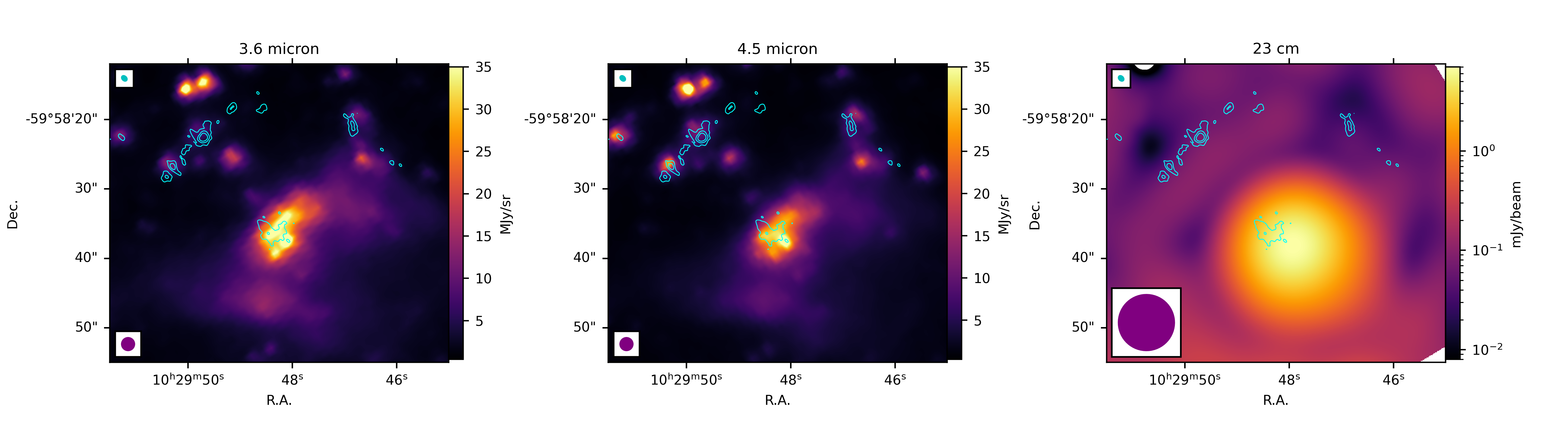}
 \caption{Radio emission at 23\,cm from the SMGPS in colorscale (units Jy/beam), with overimposed cyan contours (4, 10, and 20 times the rms) of the ALMA Band 4 continuum emission at 150 GHz. The beam of the 23 cm emission is shown in purple in the bottom left corner. The beam of the ALMA 150\,GHz emission is shown in cyan in the upper-left corner}
    \label{allbands}
\end{figure}

\begin{figure}
    \centering
    \includegraphics[width=8.5cm]{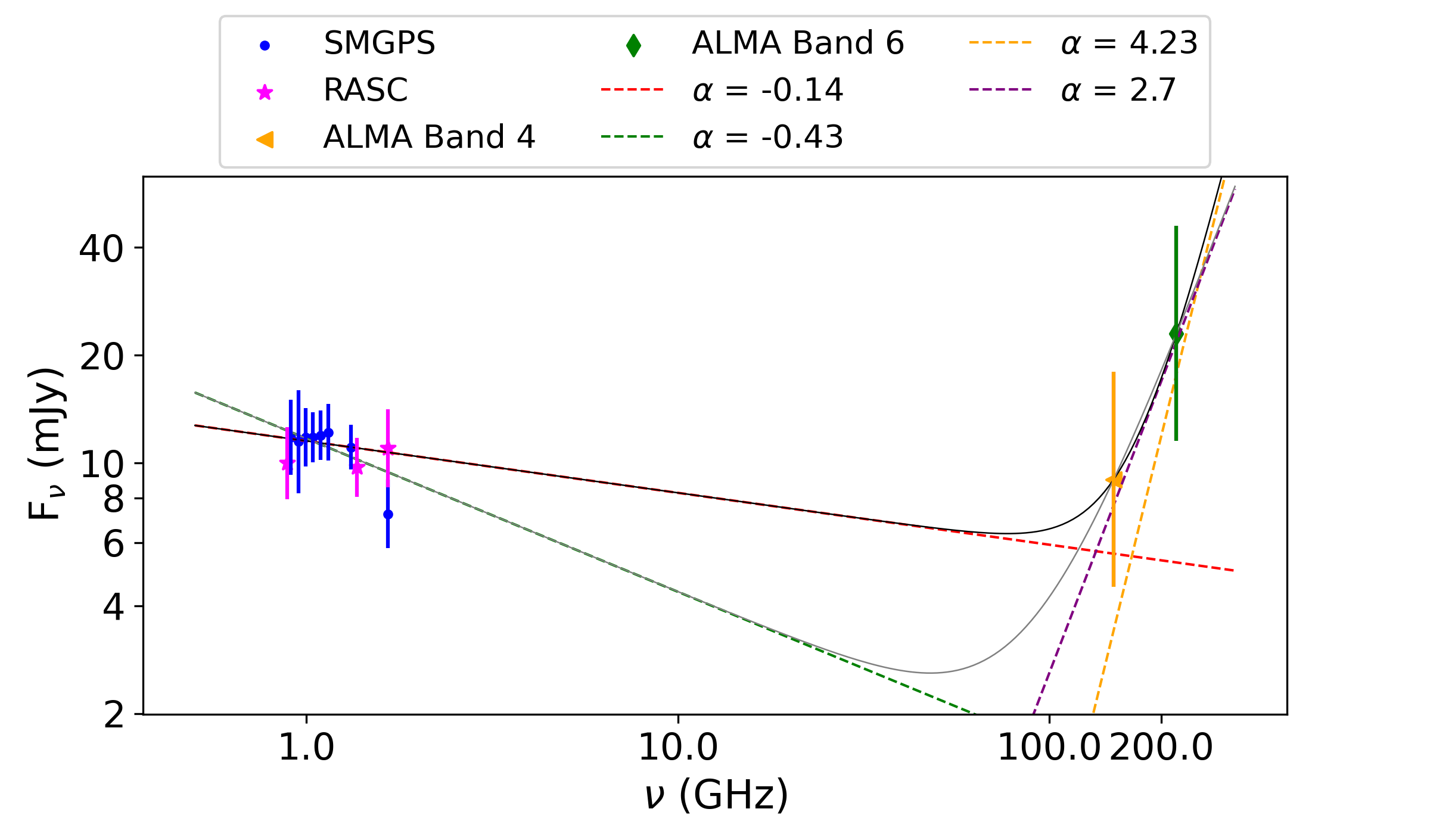}
 \caption{Spectral energy distribution of the radio emission between 0.88\,GHz and 219\,GHz from SMGPS, RACS, ALMAGAL, and the new data presented in this paper. The green dashed line is the best  linear fit between 0.88 and 1.65\,GHz including all the data points; the red dashed line is the best linear fit between 0.88 and 1.65\,GHz if the SMGPS point at 1.65\,GHz is removed; the purple dashed line is the best fit of the 150 and 219 GHz flux, after removing the free-free contamination assuming the green fit; the yellow dashed line is the same as the purple, but calculating the free-free contamination using the red dashed line.}
    \label{spectral_index}
\end{figure}
\subsection{Archival data}
We collected the continuum emission of \source\, in the infrared and at centimeter wavelength from surveys in the literature. We obtained the emission at 3.6 and 4.5$\,\mu$m from the Deep GLIMPSE survey observed by \textit{Spitzer} Infrared Array Camera (IRAC) \citep{deepGlimpse} with $\sim2$" resolution. No data for the coordinates of source \source\, were available at 5.8 and 8.0$\,\mu$m from the NASA/IPAC Infrared Science Archive (IRSA).  
We recover the emission at 23\,cm for this source from both the SARAO MeerKAT Galactic Plane Survey (SMGPS; \citealt{goedhart2024})\footnote{https://archive-gw-1.kat.ac.za/public/repository/10.48479/3wfd-e270/index.html} with an angular resolution of 8" and the Rapid ASKAP Continuum Survey (RACS; \citealt{rascI,rasc0.880GHz,RASC1.3GHz,RASC1.7GHz} ) with angular resolution of 24" at 0.88 GHz and $\sim10"$ at 1.35 and 1.65 GHz. 
\section{Analysis}

\subsection{Millimeter continuum emission}
In Figure \ref{fig:3continuum} we show 150\,GHz continuum emission from the ALMA Band 4 data (\textit{left}) and the 219\,GHz continuum emission from the ALMAGAL \textit{\sc 7m+tm2} and \textit{\sc 7m+tm2+tm1} data (\textit{center} and \textit{right}, respectively). \cite{ALMAGAL3} and Coletta et al. in prep. used the algorithm CuTEx \citep{cutexmolinari2011,cutexsoftware} to identify the compact sources inside the ALMAGAL continuum \textsc{ 7m+tm2} and \textsc{ 7m+tm2+tm1} images.
CuTEx identifies compact sources as maxima of curvature in maps by analyzing their second derivatives. This method allows an easier identification of compact sources, which stand out in the curvature images, while diffuse and background emission is strongly dampened. Once the peaks  in the curvature maps are identified, they are fitted (in the original map) with 2D elliptical Gaussians plus a background that can be of first  
or second order polynomial\footnote{The best order for the background is automatically assessed by CuTEx.}, in order to obtain the positions
and sizes of the sources (semi-major axis $\theta_A$, semi-minor axis $\theta_B$, and position angle) and the integrated flux (background subtracted). 
We use the identification numbers defined by CuTEx on the ALMAGAL continuum \textsc{ 7m+tm2}, already shown in \ref{fig:3continuum}, to identify the various sources throughout this work. CuTEx identified six sources at $\sim 7600$\,au resolution, but only four at $\sim 1400$\,au resolution \citep{ALMAGAL3}, three of which are subfragments of the source \#4 in the 7600\,au data. Therefore, there are four compact sources identified in the  7600\,au resolution image that have no counterparts in the 1400\,au resolution image in the 219\,GHz continuum data. These are core \#1, \#2, \#5, and \#6. These sources are prestellar core candidates, without a centrally peaked structure yet, as it will be discussed in Sect. \ref{sec:radialprofile}. This is supported by the fact that the integrated fluxes of cores \#2,\#5, and \#6 are larger than or comparable to the integrated flux density of core \#3 (see Table \ref{tab:massvirial}).\\
\indent The 150 GHz continuum data (\textit{left panel}) shows an emission overall similar
to that of the 219 GHz \textsc{7m+tm2} data. However, the emission
of core \#1 is not as bright as the rest of the cores and it is not
compact, while from the CuTEx algorithm extraction we performed
on this continuum image the compact core \#5 is resolved
into two cores, due to the better angular resolution. Moreover,
this new dataset has a larger FOV than the ALMAGAL data,
which reveals the presence of further emission. Considering this
newly observed area it is evident that the compact cores are located
on an arch shaped region centered close to the position of
core \#1. The radius of this arch is $\sim$18.0" which corresponds
to $\sim$0.75\,pc at the distance of \source. There are
three new cores identified along the arch outside the ALMAGAL
FOV, which are been labeled as core \#7, \#8, and \#9, respectively
(see Fig. \ref{fig:3continuum}). In Table \ref{tab:massvirial} we report the coordinates, the integrated flux densities of the cores, and their mean radius r, as derived from CuTEx
($r=d \tan\theta$, where $\theta=\sqrt{\theta_A \theta_B}$ and $d=8.5\,$kpc is the source distance) of cores from \#2 to \#9. \\ \indent
In the last column of Table \ref{tab:massvirial} we report the average column density of each core. All the cores show similar values, therefore the "disappearing" nature of some of them is not related to their average density. Moreover, this ensures that the molecular emission that we will discuss in Sect. 3.3 is not biased (e.g. the density of the core is not above the critical density of the line observed) by the different density in different cores.
\begin{figure*}
    \centering
    \includegraphics[width=18.0cm, trim={4.5cm, 8cm, 0cm, 2cm}, clip=True]{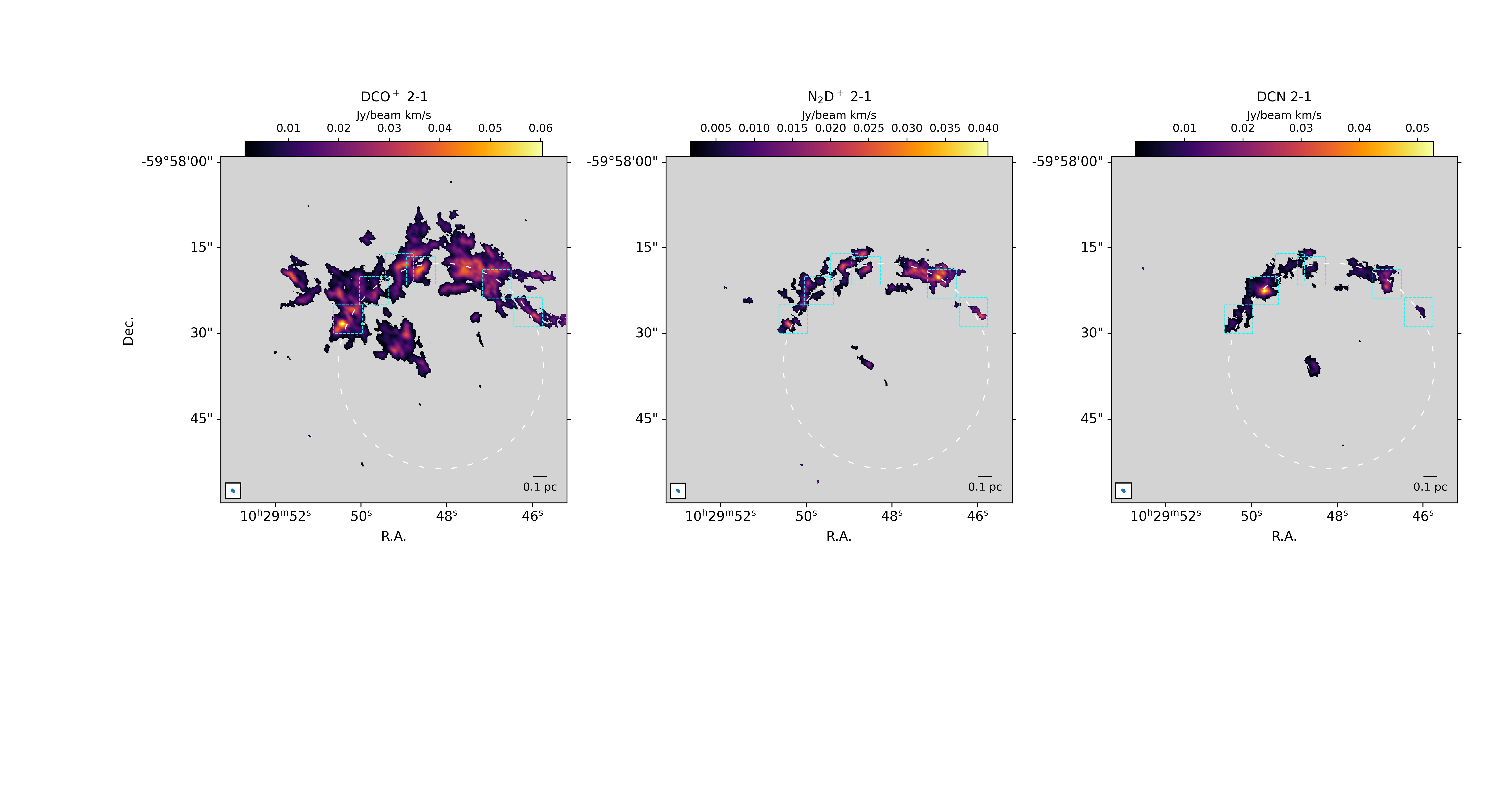}\\
     \includegraphics[width=18.0cm, trim={0cm, 0cm, 0cm, 1cm}, clip=True]{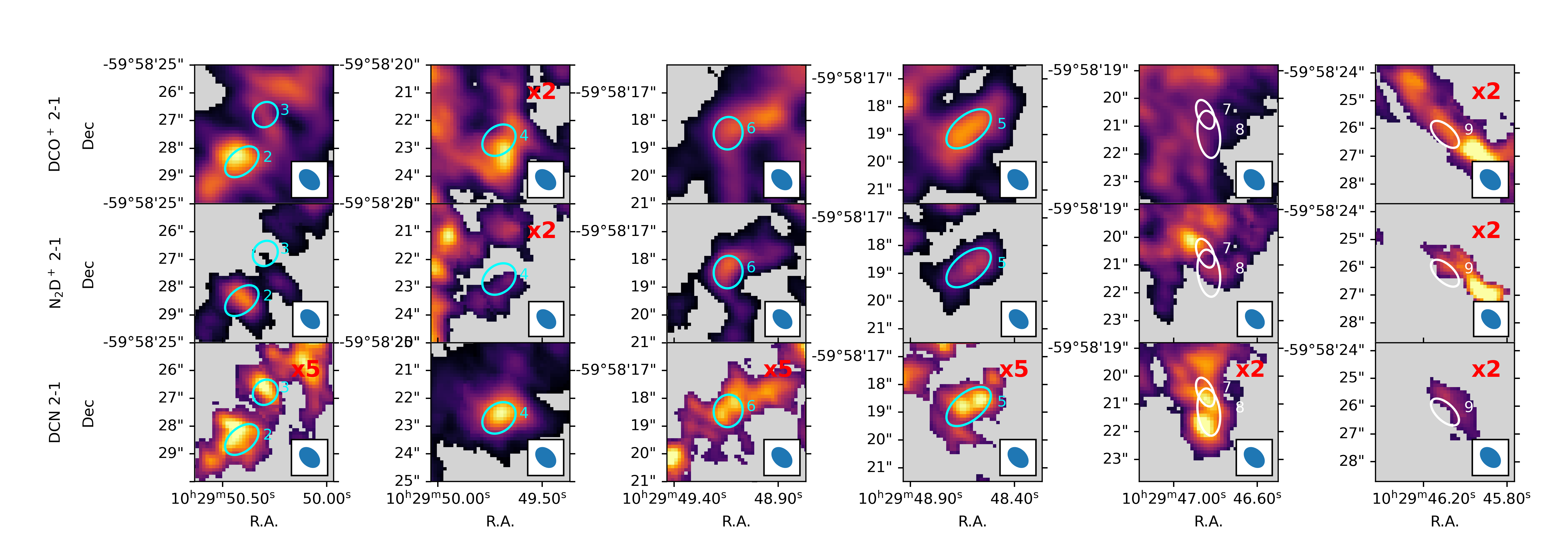}
         \caption{Top panel: moment-0 map of \dco (2$-$1) (left), \ntwod (2$-$1) (middle), and DCN (2$-$1) (right); bottom panels: zooms-in of the three moment-0 maps in the boxes delimited in cyan in the upper panels. The cyan and white ellipses identify the compact cores extracted with CuTEx from the ALMAGAL \textsc{7m+tm2} continuum and ALMA Band 4 continuum, respectively. Some of the zoom-in boxes have the intensity multiplied by a factor reported in red in the top-righ corner of the box, to highlight the emission with respect to the colorscale adopted for the full maps. }
    \label{Mom0}
\end{figure*}

\begin{table*}[]

    \centering
    \caption{Properties of the continuum cores along the arch shaped structure.}
    \begin{tabular}{cccccccccc}
    \hline\hline\\
        ID & l & b & $F_{\rm{INT}}$ & $r$ & $T$ & $M$ & $FWHM$ & $\alpha_{\mathrm{vir}}$  & $n$(H$_2$) \\
        & [$^{\circ}$]&[$^{\circ}$] & \small{[mJy]}  & \small{[10$^{3}$ AU]} & \small{[K]} & \small{[M$_{\odot}$]} &\small{[km\,s$^{-1}$]} & &\small{[10$^6$ cm$^{-3}$]} \\ \hline
        \multicolumn{10}{c}{\textit{\tiny{cores identified in ALMAGAL 7MTM2 continuum: disappearing}}}\\
        2& 286.07362& -1.821405 &1.8  &4.7 &10-20 & 3.1-8.3 &1.2 &0.8-2.2 & 0.9-2.4\\
        5&286.06904  &-1.821022 &3.6  &5.9 &10-20 &6.2-16.6 &0.97 &0.3-0.9 &0.9-2.4 \\
        6&286.06992  &-1.820381 &2.6  &4.7 &10-20 &4.5-12.2 &1.1 &0.5-1.3 &1.3-3.5 \\
        \hline
        \multicolumn{10}{c}{\textit{\tiny{cores identified in ALMAGAL 7MTM2 continuum: not disappearing}}}\\
        3& 286.07318 & -1.821122 &1.9  &3.8 &20-30 &1.9-3.2 &1.2&1.7-2.8  & 1.0-1.8\\
        4&286.07154 &-1.820783 &10.6  &4.9 &$39\pm5$ & 8.2 &1.8 &2.0 & 2.1 \\

        \hline
        \multicolumn{10}{c}{\textit{\tiny{new cores identified in ALMA Band 4 continuum}}}\\
        7 & 286.06613 & -1.823351 &0.48 & 3.5 & 20-30 & 1.9-3.1 & 1.7 & 3.3-5.3 & 1.4-2.2  \\ 
        8 & 286.06620 & -1.823540 & 1.5 & 5.0 & 20-30 & 6.1-9.7 & 1.7 & 1.5-2.4 & 1.5-2.4   \\ 
        9 & 286.06559 & -1.825501 & 0.9 & 3.8 & 20-30 & 3.6-5.8 & 0.7 & 0.3-0.5 & 3.2-2.0  \\ 
                   \hline\hline
    \end{tabular}
    \flushleft
    {\small{\textbf{Notes:} the integrated flux density $F_{\rm{INT}}$ for cores \#2 to \#6  is at 219 GHz, while for the remaining cores is at 150 GHz. The FWHM is the mean value of those of detected species on the cores.}}
    \label{tab:massvirial}
\end{table*}
\subsection{Centimeter continuum emission}
In Fig. \ref{allbands} we show the continuum emission of this region at  23\,cm from SMGPS, with superimposed the contours of continuum emission from ALMA observations at 150\,GHz. The center of the arched emission is consistent with the position of bright radio emission. The position of the peak of the 23\,cm emission matches within 3" with the position of a candidate \ion{H}{ii} region from the catalog of \citet{anderson2014}, which used the morphology seen in the emission in WISE bands to identify \ion{H}{ii} regions, together with radio emission and line emission. The source was classified as \textit{candidate} due to a lack of data from radio recombination lines (RRL) to confirm the presence of ionized material. \\ \indent  
We collected all the available archival data in the literature to build the most complete radio spectral energy distribution (SED) of this source, that encompass core \#1 identified in the continuum images of ALMA Band 4 and 6.
Using the data-cube with the emission in different sub-bands for SMGPS, and the emission at the three wavelengths of RACS we were able to construct the SED between 0.88\,GHz and 1.65\,GHz. The source is point-like, as a preliminary 2D Gaussian fit performed in Cube Analysis and Rendering Tool for Astronomy (CARTA, \citealt{CARTA}) matches the angular resolution of the data. We then perform aperture photometry using a circular aperture of 2 times the resolution of the data ($\sim$20").
The SED in the available range shows a flat profile (see Fig. \ref{spectral_index}).  The spectral index $\alpha$ (assuming $S_{\nu}\propto\nu^{\alpha}$) from the fit of the data is $-0.4\pm0.4$, mostly driven by the point at 1.65\,GHz from SMGPS, while the flux density from RACS data at the same frequency is consistent with a $-0.1\pm0.4$ spectral index. The spectral index values and their error have been derived with a Monte Carlo Markov Chain linear fit to the points between 0.8\,GHz and 1.6\,GHz, using 100 walkers and 500 steps. Both values of $\alpha$ are consistent with optically thin free-free emission, as seen from $\alpha$ values derived in the same range from known \ion{H}{ii} regions using SMGPS from \citet{goedhart2024}. We included in the SED also the flux densities inside the region of emission at 23\,cm in the two ALMA Bands that are dominated by the dust emission, which correspond to what we previously defined as core \#1. We calculated the spectral index between the point at 150\,GHz and at 220\,GHz, first removing the contribution from free-free emission extrapolated from the two possible fits of the first part of the SED. The range of the spectral index is then between 2.7 (assuming $\alpha =$-$0.43$ for the free-free emission) and 4.2 (assuming  $\alpha =$-$0.14$ for the free-free emission), which corresponds to the emission of dust along the same line-of-sight of the emission of the \ion{H}{ii} region. The value of the exponent of the power-law of dust emissivity $\beta$ (with $\beta=\alpha-2$ in the Rayleigh-Jeans regime) for this source is in the range of 0.7-2.2. These values are consistent with typical values found in the ISM \citep{Planck_abergel2014}. The commonly used value of $\beta\sim1.7-2$ corresponds to grain distribution with sizes between 100\,\textup{~\AA} and 0.3\,$\mu$m \citep{Weingartner2001}, while larger grains result in lower values of $\beta$, reaching values below 1 for grain sizes above 1\,mm \citep[][and references therein]{cacciapuoti2023}. However, given the more diffuse morphology of core \#1 from its 150\,GHz emission, we do not consider it to be a compact core and we exclude it from further analysis to determine whether this is a prestellar core.  \\ 
\indent The number of Lyman continuum photons emitted by the ionizing source and that generate the \ion{H}{ii} region can be calculated using the radio
flux density, following the relation given by \citet{matsakis1976}: 
\begin{equation}
    N_{Ly } = 7.5\times10^{46}\,\Biggl(\frac{S_{\nu}}{\rm{Jy}}\Biggr)\,\Biggl(\frac{D}{\rm{kpc}}\Biggr)^{2}\,\Biggl(\frac{\nu}{\rm{GHz}}\Biggr)^{0.1}\,\Biggl(\frac{T_e}{\rm{10^4\,K}}\Biggr)^{-0.45}\, \mathrm{phot.\,s^{-1}}\,,
\end{equation}
where $S_\nu$ is the flux density of the source at frequency $\nu$, $D$ is the distance of the source, and $T_e$ is the electron temperature. \citet{balser2011} provided a relation between $T_e$ and the galactocentric distance of a source, that estimated at the galactocentric distance of \source\,\, ($\sim10\,$kpc) gives a value of $T_e\sim 9000\,$K. Given the presence of some scatter in the relation, we considered a range of $T_e=8000-10000\,$K.  Considering the flux density from the average continuum map of SMGPS to be 10.5\,mJy at 1.36\,GHz, and using the source distance of 8.5\,kpc the rate of ionizing photon is in the range $5.9-6.5\times10^{46}$\,photons\,s$^{-1}$, corresponding to a star of spectral type B0.5 \citep{panagia1973}.

\begin{table*}
    \caption{Molecular emission from the continuum cores along the arch shaped structure.}

    \centering
    \begin{tabular}{ccccccccccccc}
    \hline\hline\\
 & \multicolumn{3}{c}{\dco (2--1)} & \multicolumn{3}{c}{\ntwod (2--1)} & \multicolumn{3}{c}{DCN (2--1)} & \multicolumn{3}{c}{CH$_3$CCH (9--8)} \\
\cmidrule(lr){2-4}
\cmidrule(lr){5-7}
\cmidrule(lr){8-10}
\cmidrule(lr){11-13}
ID & det  & peak & FWHM  & det  & peak & FWHM & det  & peak & FWHM & det  & peak &FWHM \\
& & & {\small [km/s]} & & & {\small [km/s]} & & & {\small [km/s]}& & & {\small [km/s]}\\ 
\hline
        \multicolumn{13}{c}{\textit{\tiny{cores identified in ALMAGAL 7MTM2 continuum: disappearing}}}\\
        2 & Y & Y & $1.11\pm0.04$  & Y & Y&$1.22\pm0.07$ & Y & Y&$1.3\pm0.1$ & N & N&- \\
        5 & Y & Y & $0.97\pm0.04$  & Y & Y&$1.01\pm0.05$ & Y & Y&$0.93\pm0.07$ & N & N&-\\
        6 & Y & Y & $0.97\pm0.05$  & Y & Y&$1.07\pm0.06$ & Y & Y&$1.09\pm0.08$ & N & N&- \\
        \hline%
        \multicolumn{13}{c}{\textit{\tiny{cores identified in ALMAGAL 7MTM2 continuum: not disappearing}}}\\
        3 & Y & N &$1.13\pm0.06$ & N & N &- & Y & Y&$1.23\pm0.12$ & N & N       &- \\
        4 & Y & N &$1.53\pm0.13$   &N & N &- & Y & Y&$2.1\pm0.12$ & Y & Y& $1.9\pm0.1$  \\
        \hline
        \multicolumn{13}{c}{\textit{\tiny{cores identified in ALMA Band 4 continuum outside the ALMAGAL FOV}}}\\
        7 & Y &N &$1.73\pm0.12$ & Y & N&$1.8\pm0.3$ & Y & Y&$1.66\pm0.16$ & N & N&- \\
        8 & Y &N &$1.58\pm0.13$ & Y & N&$1.7\pm0.3$ & Y & Y&$1.67\pm0.12$ & N & N&- \\
        9 & Y &N &$0.70\pm0.06$& N & N&- & N & N&- & N & N&-    \\
                   \hline\hline
    \end{tabular}
    \flushleft
    {\small{\textbf{Notes}: det: Y if the transition is detected at S/N>5 in the spectra averaged on the core; peak: Y if the emission from the moment-0 map has a peaked morphology inside the area of the core. We do not took into consideration the hyperfine splitting of the deuterated species, so the FWHM values reported are upper limits.}}
    \label{tab:moleculardetection}
\end{table*}

\begin{figure*}
    \centering
    \includegraphics[width=6.5cm, trim={4.5cm, 8cm, 30cm, 3.cm}, clip=True]{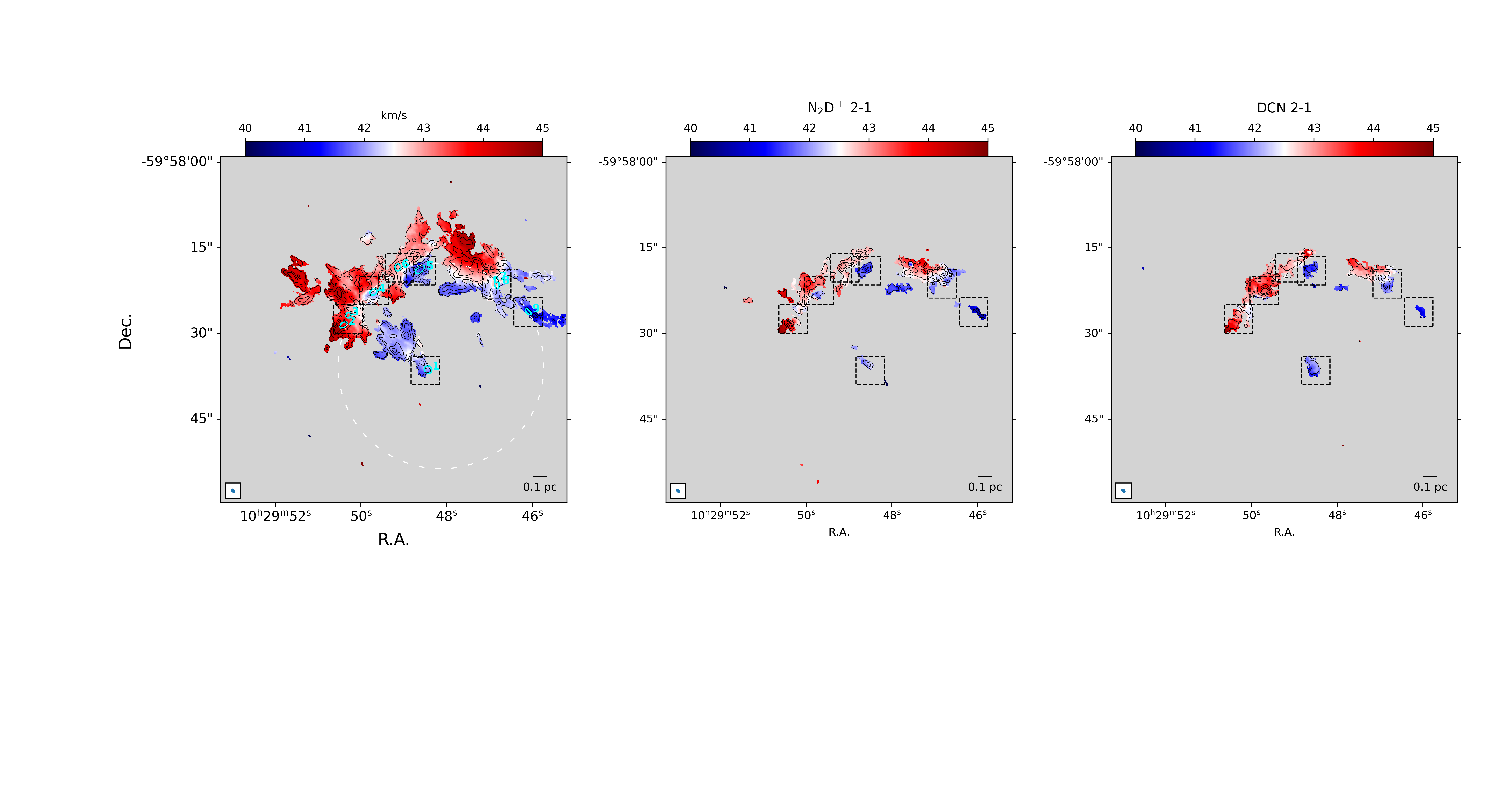}
    \includegraphics[width=6.5cm, trim={4.5cm, 8cm, 30cm, 3.cm}, clip=True]{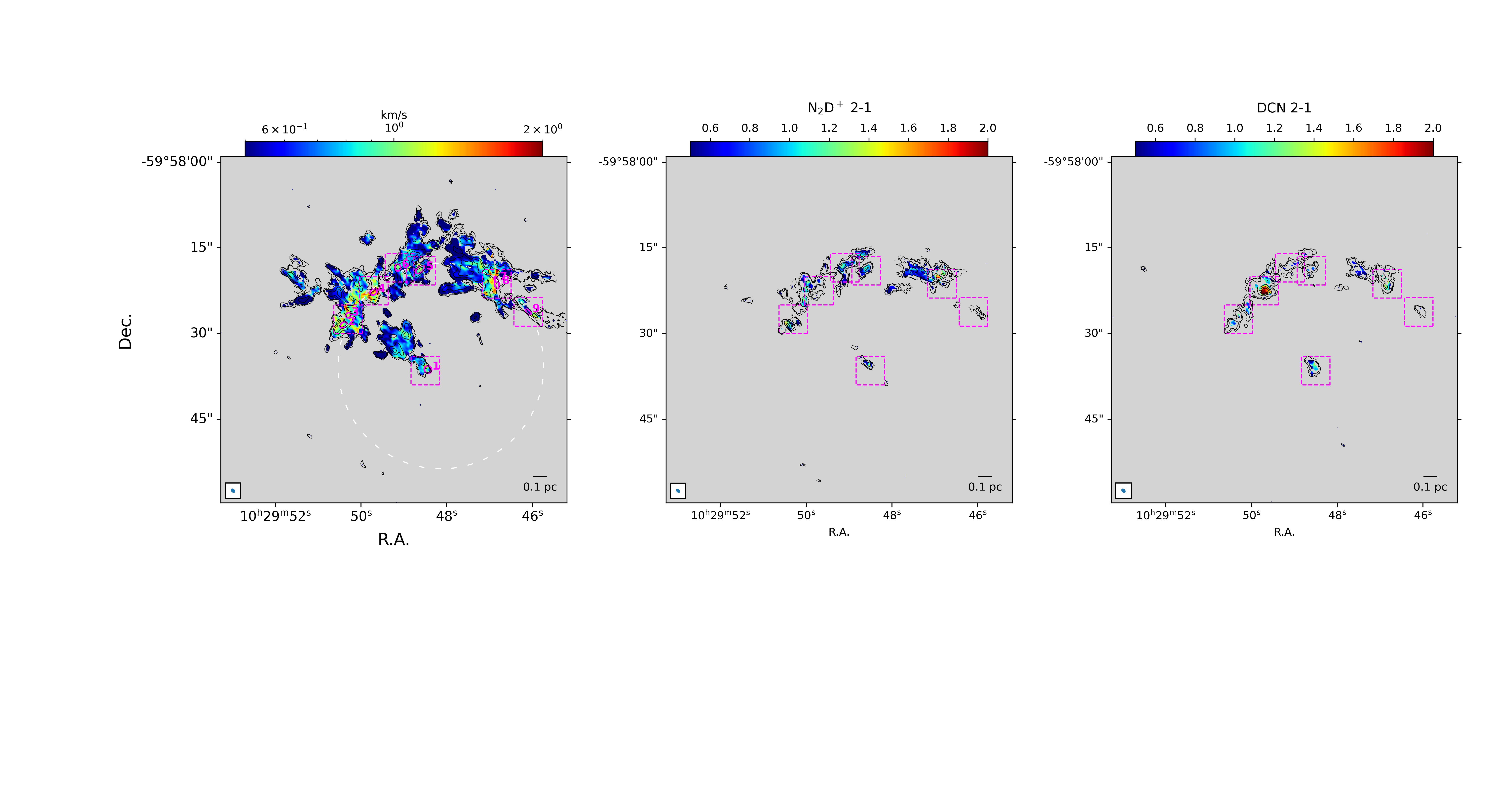}
    \caption{Moment-1 (left) of \dco (2$-$1) and FWHM (right) from the fit pixel-by-pixel of the line profile of \dco (2$-$1). The color scales unit is km\,s$^{-1}$. The dashed line boxes are the same as in Fig. 1. The cyan ellipses are the compact cores detected. The white dashed circle is the 0.75\,pc radius circle along which the cores are aligned. }
    \label{Mom1-2}
\end{figure*}
\subsection{Molecular emission}
\begin{figure*}
    \centering
    \includegraphics[width=19cm]{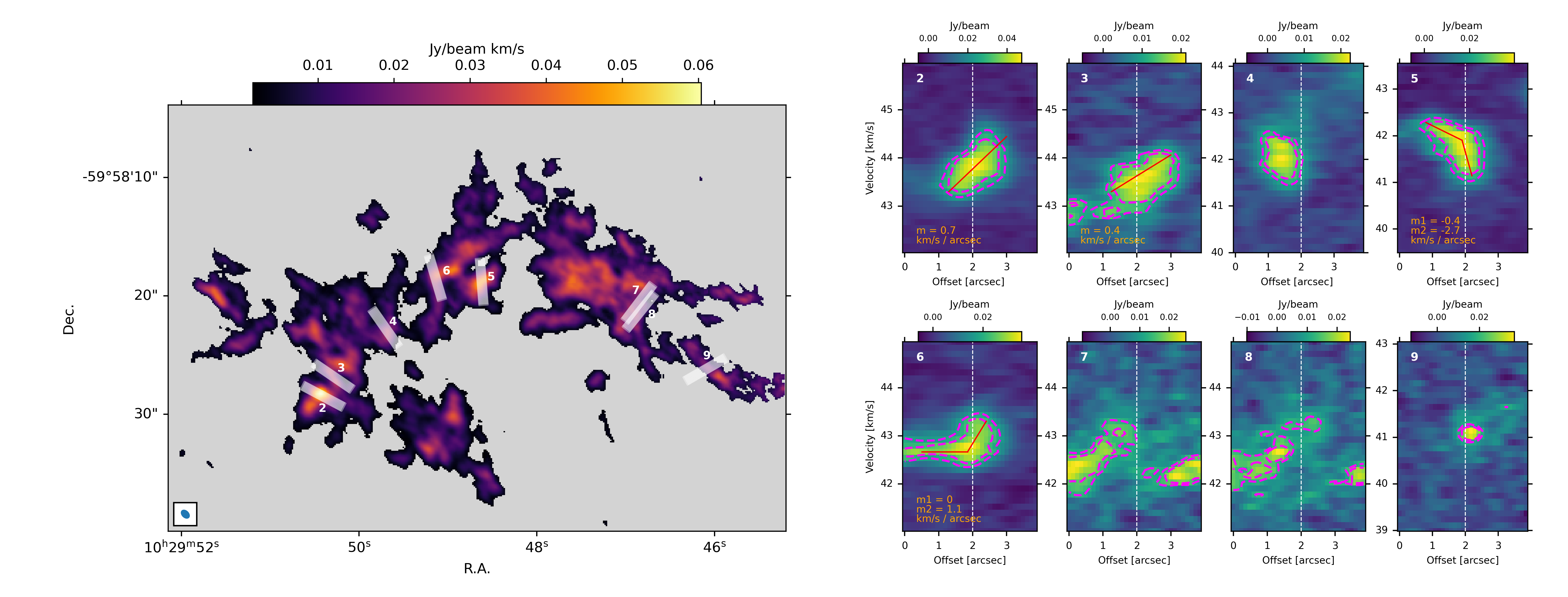}
    \caption{Left panel: moment-0 maps of \dco (2$-$1) with shown in white the cuts for the position-velocity plots, centered at the peak positions of the cores. Right panels: position-velocity plots for each core in the arch. The white vertical line delimits the position of the peak of the core. The magenta dashed lines delimits the contours where the emission is half and two thirds of the maximum.}
    \label{pvplot}
\end{figure*}

\begin{figure}
    \centering
    \includegraphics[width=\linewidth]{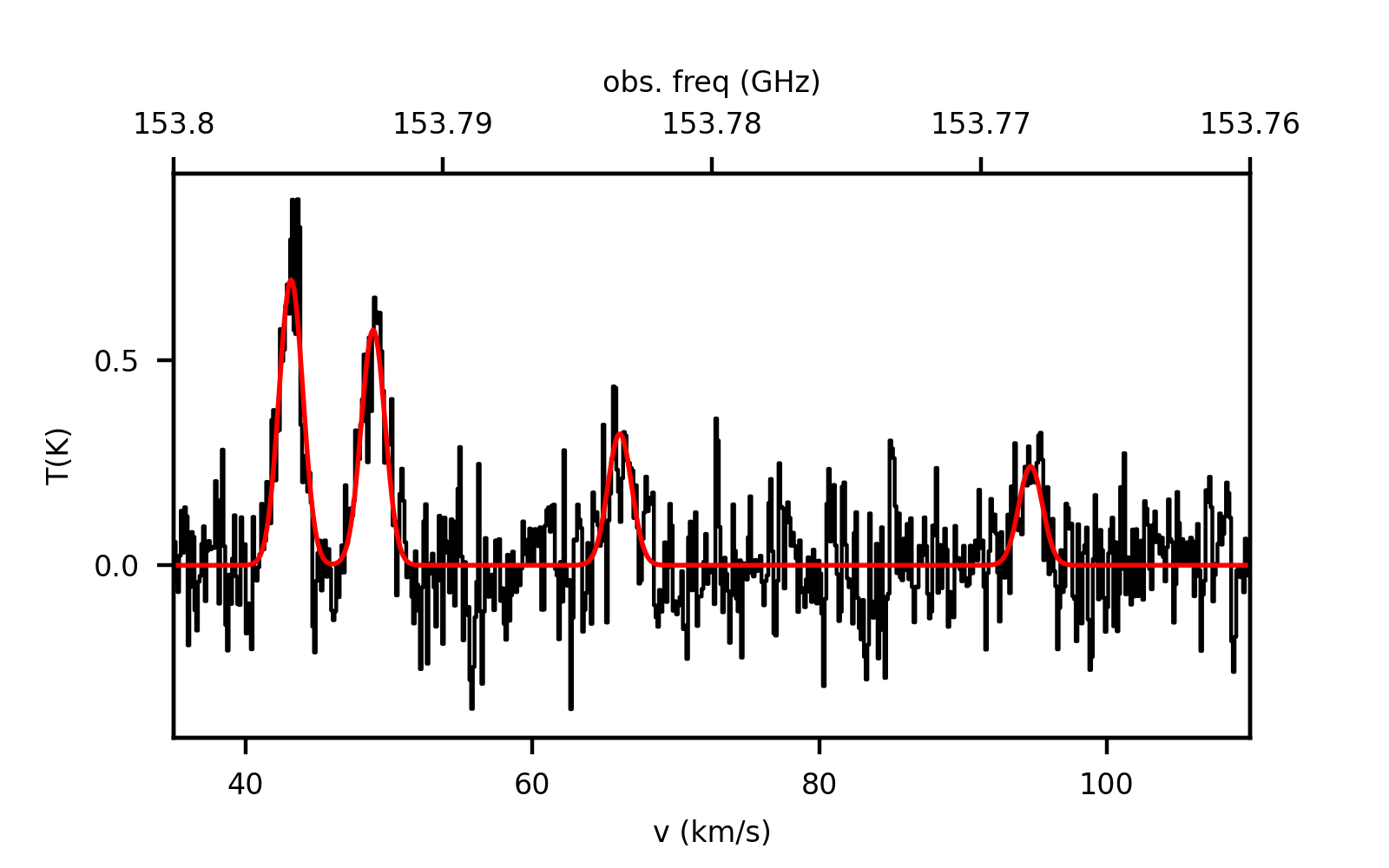}
    \caption{Average spectrum of CH$_3$CCH (9-8) over core \#4 in black. In red we show the simulated spectrum obtained using the best-fit parameters with MADCUBA.}
    \label{fig:ch3cch}
\end{figure}

Figure \ref{Mom0} shows the integrated intensity (moment-0) maps of \dco\,(2$-$1) (\textit{upper left}), \ntwod\,(2$-$1) (\textit{upper center}), and DCN\,(2$-$1) (\textit{upper right}) emission. The moments have been created using a velocity range defined pixel-by-pixel. We created the mask used for the moments integration following these steps: 1) we calculated the root mean squared ($\sigma$) of the cube for each pixel, using the first 500 channels of the cubes, which were determined to be free of any emission; 2) we created a first mask of the cube to include the voxels of emission above 5$\sigma$; 3) we created a second mask to include the voxels of two consecutive channels with emission above 2$\sigma$; 4) we eliminated from the mask in step 3 all the voxels that do not include a previous voxel above 5$\sigma$, i.e. defined in the mask in point 2; 5) we enlarge the mask obtained from point 4 of $0.25\times$FWHM of the beam in the spatial directions and of two channels at the beginning and at the end of the range of integration in the spectral direction, to include possible faint emission at the tail of the line profile. The mask obtained in step 5 is the final mask that we used to create the moments.  As a last step, from the spectral rms and the number of channels of integration of each pixel, we constructed a map of the moment-0 rms, and blanked from the moment maps the pixels where the moment-0 signal-to-noise ratio (S/N) was below 3.\\ \indent
Moreover, we extracted averaged spectra towards the cores (see Figs. \ref{fig:detection} and \ref{fig:detection2}) and we report the detection at S/N>5 of the transitions observed in ALMA Band 4 in Table \ref{tab:moleculardetection}. From Fig. \ref{Mom0}, the emission of \dco (\textit{top row of the lower panel}) is bright and compact towards the disappearing cores \#2, \#5, and \#6, while even if still present does not peak or shows enhancement towards cores \#3, \#4, \#7, \#8, and \#9. An enhancement and peak morphology is seen for \ntwod (\textit{middle row of the lower panel}) towards cores \#2, \#5, and \#6; in the remaining cores it is detected from the spectra and moment-0 maps only towards cores \#7 and \#8, but without a peaked morphology. Conversely, the emission of DCN (\textit{bottom row of the lower panel}) shows a bright peaked emission towards the position of core \#4. However, DCN is  detected toward the majority of other cores. Its morphology always shows a peak inside the area of the cores, often offset from the center of the cores, or with two peaks.\\
\indent As for the continuum, the molecular emission in ALMA Band 4 revealed the presence of  material outside the continuum cores with other spots in which the emission is enhanced in those molecular species both outside and inside the ALMAGAL FOV. Those could also be the positions of further prestellar cores, better revealed by the molecular emission, as seen in two star-forming regions with H$_2$D$^+$ by \citealt{redaelli2021}. Table \ref{tab:moleculardetection} gives a summary of where the different molecular species shows a peaked emission.\\ \indent  We also report in Fig. \ref{Mom1-2} the velocity of the gas along the line of sight and the full-width at half maximum (FWHM) of the line using \dco (2$-$1) which shows the most widespread emission in the region.
In the moment-1 map we identify a velocity gradient along the radial direction centered on the \ion{H}{ii} region position, signature of an expanding shell. Similar gradients have been found in other shell around ultra compact \ion{H}{ii} (UC\,\ion{H}{ii}) regions  and \ion{H}{ii} regions \citep[e.g.][]{fontani2012,wang2016}.  Cores are located at the position where the velocities match the median of the velocity gradient, with the exception of core \#1. The FWHM of the \dco\,(2$-$1) is narrow, with values between 0.6 and 1.5 km\,s$^{-1}$ in the largest part of the region, with the exception of core \#4 where it reaches values up to $\sim2$\,km\,s$^{-1}$. \\ \indent 
To further investigate the presence of an expanding shell or collision between the medium outside the shell and the ionizing front that potentially triggered new star formation, we report in Fig. \ref{pvplot} the position-velocity (PV) plots of the \dco (2$-$1) emission along cuts at the positions of the cores in the radial direction, i.e. along the direction that connects each core with the center of the \ion{H}{ii} region. For cores \#4, \#7, \#8, and \#9 there is not bright enough emission coincident with the center of the cut, i.e. the peak position of the core, as already seen in the moment-0 map. For the other cores we observe a radial velocity gradient, which in the case of cores \#5 and \#6 has a breaking point at their position. As stated above, the radial velocity gradient is likely the signature of an expanding shell, while the breaks in the gradient in the p-v plot resemble those showed in Fig. 5 of \cite{arce2011} at the interface between the expanding bubble and the material surrounding it.
\\ \indent The CH$_3$CCH (9-8) emission has only been detected towards core \#4, with the transitions $K=0$, and 1 clearly detected above 5 sigma, and two other transitions marginally detected. We extracted an average spectrum towards this core, and used the Spectral Line Identification and Modelling (SLIM) tool of the MAdrid Data CUBe Analysis package (MADCUBA \citealt{MADCUBA}) to perform a line fitting. 
To analyze CH$_3$CCH we selected the Cologne Database for Molecular Spectroscopy catalog (CDMS, \citealt{cdms})\footnote{The entry of CDMS for CH$_3$CCH is based on the following works: \cite{ch3cch_1,ch3cch_2,ch3cch_3,ch3cch_4}.}, and used SLIM to generate synthetic spectra under the assumption of local thermodynamic equilibrium (LTE). SLIM allows to fit the spectrum using non-linear least-squares LTE fit to the data using the Levenberg–Marquardt algorithm, comparing the modeled spectra with the observed one. The free parameters fitted by the LTE model
are the molecular column density ($N$), excitation temperature ($T_{\rm{ex}}$), central velocity of the line emission, and FWHM of the Gaussian line profiles. 
The resulting best fit to the spectrum is shown in Fig. \ref{fig:ch3cch}). The derived temperature is of $T_{\rm{ex}}=39\pm5\,$K, while the column density of this species is $N=(2.1\pm 0.2)\times10^{14}$\,cm$^{-2}$.\\ 
\indent Lastly, to investigate the possible presence of outflows, we created the moment-0 and moment-1 maps of SiO $(5-4)$ transition, using the \textsc{7m+tm2} cubes of the ALMAGAL project. These are shown in Appendix in Fig. \ref{fig:SiO}. Among the cores inside the ALMAGAL FOV, only the core \#4 is associated to SiO emission, which shows a velocity gradient in the North East-South West direction.

\section{Discussion}
\subsection{Radial profile of continuum emission}
\label{sec:radialprofile}
\begin{figure}
    \centering
    \includegraphics[width=0.98\linewidth]{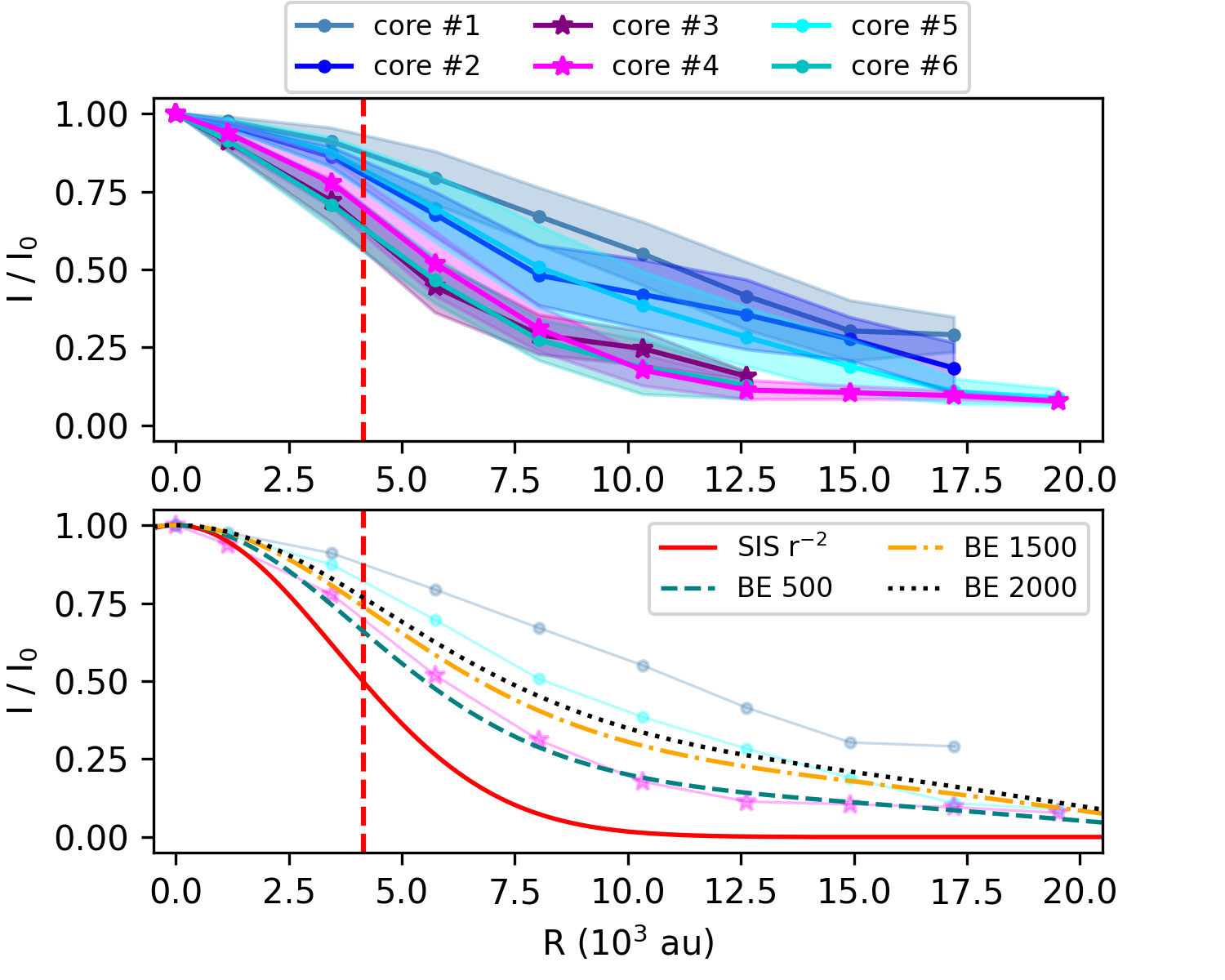}
    \caption{Upper panel: radial profile of the emission of cores \#1 to \#6 in the ALMA 219\,GHz \textsc{7m+tm2} continuum images. Lower panel: radial profile of the emission expected from isothermal cores with known radial density distribution observed with the resolution of ALMA 219\,GHz \textsc{7m+tm2} data. We report the observed profiles of cores \#1, \#4, and \#5 for visual comparison. The dashed vertical red line indicates the beam half width at half maximum of the 219\,GHz continuum data. }
    \label{fig:radialprof}
\end{figure}

We used the continuum emission in ALMA Band 6 \textit{\sc 7m+tm2} data to construct the radial profile of emission of the cores from \#1 to \#6, and then compare the profiles of disappearing and not-disappearing ones. For every core from \#1 to \#6 we cut a region around the core in which by visual evaluation the emission is not affected by the presence of nearby cores. Each core has been identified as an ellipse with a specific center position (x$_0$, y$_0$),  position angle (PA), semi-major to semi-minor axis ratio ($\theta_{A}/\theta_{B}$), and semi-major axis from the algorithm CuTEx. Using the four parameters x$_0$, y$_0$, PA, and $\theta_{A}/\theta_{B}$ fixed to the value defined by CuTEx for the core, it is possible to assign to each pixel of the core the semi-major axis (that we refer to as the profiles radius) of the ellipse passing through that pixel. We then create bins of distance and obtain the mean value of intensity and the standard deviation in the bin. The profiles obtained are shown in the upper panel of Fig \ref{fig:radialprof}. \\\indent The radial profile of all the disappearing cores except core \#6 are shallower than those of cores \#3 and \#4, which have been detected also in the ALMAGAL \textit{\sc{7m+tm2+tm1}} data. This confirms our initial guess that the disappearing cores have a less peaked density profile. \\ \indent To further characterize this difference, in the lower panel of the same figure, we plot the radial profile of emission of known profile density distributions, alongside the profiles of core \#1, \#4, and \#5 for reference. Specifically, we plotted the emission from the radial profile of a singular isothermal sphere \citep[SIS,][]{shu1987},  $\rho(r)\propto r^{-2}$, and of a Bonnor-Ebert sphere \citep[BE,][]{ebert1955Z, Bonnor1956} characterized by a central flat profile up to a radius close to the critical value $R_0$ and by a decrease with $r^{-2}$ for $r\gg R_0$. We approximated the Bonnor-Ebert sphere using the analytic radial profile $\rho(r)\propto (1+(r/R_0)^{\alpha})^{-1}$, with $\alpha=2$ (see e.g. \citealt{fischera2014}) and using three different values of $R_0$: 500\,au, 1500\,au, and 2000\,au. Assuming that the cores are isothermal and that the dust properties are not varying with the radius (i.e. $\kappa_{\nu}(r)=\rm{const.}$), the radial profile of the flux density emitted is given by the convolution of the surface density profile $\Sigma(R)$ with a gaussian profile with FWHM the size of the beam of the ALMA Band6  \textit{\sc 7m+tm2} data, where $R$ is the radial coordinate on the plane of the sky. $\Sigma(R)$  is derived integrating the density profile $\rho(r)$ along the line of sight (x direction, $r^2 = R^2+x^2$ ):
\begin{equation}
    \Sigma(R) \propto \int_{-\infty}^{+\infty}{\rho(\sqrt{x^2+R^2})dx} \propto \int_{R}^{+\infty}{\frac{2\rho(r)r}{\sqrt{r^2-R^2}}dr}\,.
\end{equation}
\indent From the visual comparison of the flux density expected by the assumed density profiles with the observed profiles, those of cores \#3, \#4, and \#6 are consistent with Bonnor-Ebert profiles with a central flattening of $\sim500\,$au, while the profiles of cores \#2 and \#5 are better reproduces by Bonnor-Ebert spheres with a central flattening over a larger area, up to a radius of 1500-2000\,au. Values up to $\sim1500\,$au for the radius of the central flat region in prestellar cores have been already observed in the literature, e.g. in the  prestellar core L1544 \citep{caselli2019}. The profile of core \#1 is flatter than the ones expected by all these models. 

\subsection{Evolutionary stages of the cores from chemistry}

The enhancement in the abundance of \ntwod and \dco is associated with early stages of star formation, 
where the freeze-out of the CO onto dust grains at densities above $10^5\,$cm$^{-3}$ and temperatures below 20\,K removes the 
main destruction pathways of the H$_2$D$^+$ ion, and starts the deuteration process \citep{millardeuteration,caselli2012,fontanideuteration}. This is not the case for DCN which is often found to be more abundant in warmer gas, thanks to chemical pathways efficient at higher temperature 
\citep{millardeuteration, roueff2007, parise2009, gieser2021, cunningham2023DCN}.\\\indent \cite{fontanideuteration} show that deuterium fractionation from 
N$_2$H$^+$ is extremely sensitive to gas  temperature, and high-mass starless cores with warmer temperatures (above 20\,K) have a drop 
of one order of magnitude in the D/H ratio. The extreme dependence on gas temperature is also evident in the models presented in \cite{fontani2014timescale}, 
which shows that after the temperature increases above 20\,K the D/H ratio drops after just $\sim10^2$ yrs for N$_2$H$^+$ and after $\sim10^4$ yrs for HNC. The timescale of the drop of the D/H ratio for \dco is similar to that of \ntwod ($\sim100$\,yrs), as it is possible to see in Fig. \ref{dcodrop} in Appendix. The plot shows the result of the same model run in \cite{fontani2014timescale} for a density of 10$^{5}$\,cm$^{-3}$ (private communication with K. Furuya and F. Fontani).\\
\indent We indeed found that three out of four disappearing cores (\#2, \#5, and \#6) show bright and compact emission of \dco (2$-$1) and \ntwod (2$-$1). Since the enhancement of abundance of these species due to deuterium fractionation fades quickly with the onset of star-formation and the subsequent rise in temperature above 20\,K, we confirm that these are prestellar cores in a region of clustered star-formation. This is further supported by the shallower radial profiles showed for all these cores, with the exception of core \#6. Moreover, none of these cores is associated with SiO $(5-4)$ emission, an outflow tracer, as shown in Fig. \ref{fig:SiO}.\\ \indent
The lack of peaked emission in \dco and \ntwod towards both cores \#3 and \#4, indicates that these cores have a higher temperature. Moreover, these two sources have a more peaked radial profile since they are detected in the continuum as compact sources in ALMAGAL \textsc{7m+tm2+tm1}, and as showed in Sect. \ref{sec:radialprofile}. Core \#4 also shows bright and peaked emission of DCN, whose production is more effective in warmer gas, of CH$_3$CCH, which is a complex organic molecule, also associated with more evolved stages of the star-formation process \citep{herbstvandishoeck2009, molinarich3cch, sabatini2021}, and of SiO, and outflow tracer. For this core we used the detected emission of CH$_3$CCH (9-8) to derive a temperature of $39\pm5\,$K.
To summarize the information from the molecular emission towards core \#4, \dco and \ntwod, tracers of early stages of star formation, are not enhanced towards this source, while there is bright emission of DCN, CH$_3$CCH, and SiO which emission requires more evolved stages with higher temperatures, not seen in any other core. Therefore, we conclude that core \#4 is the most chemically evolved and we assume that the temperature in the rest of the cores along the arch is lower than towards core \#4. 
The cores \#7, \#8, and \#9 do not have peaked emission in \dco (2$-$1) or \ntwod (2$-$1), but cores \#7 and \#8 have been detected with a peaked morphology in DCN (2$-$1). For these reasons, we consider them  and core \#3 as in an intermediate evolutionary stage between that of cores \#2, \#5, and \#6, and  that of the most evolved core in our sample, namely core \#4.
Based on this, we assign a plausible range of temperatures between 10\,K and 20\,K to cores \#2, \#5, and \#6, and a warmer range between 20\,K and 30\,K to cores \#3, \#7, \#8, and \#9.

\begin{table*}
\caption{Separation of the bound cores along the arch-shaped region.}
    \centering
    \begin{tabular}{cccccccccccc}
    \hline\hline\\
    \multicolumn{2}{c}{\#2-\#3} & \multicolumn{2}{c}{\#3-\#4} & \multicolumn{2}{c}{\#4-\#6}& \multicolumn{2}{c}{\#6-\#5}& \multicolumn{2}{c}{\#5-\#8}  & \multicolumn{2}{c}{\#8-\#9}\\
\cmidrule(lr){1-2}
\cmidrule(lr){3-4}
\cmidrule(lr){5-6}
\cmidrule(lr){7-8}
\cmidrule(lr){9-10}
\cmidrule(lr){11-12}

[$\arcsec$] & [pc] & [$\arcsec$] & [pc] & [$\arcsec$] & [pc] & [$\arcsec$] & [pc] & [$\arcsec$] & [pc] & [$\arcsec$] & [pc] \\
\hline
\multicolumn{12}{c}{\textit{projected}}\\
1.9 & 0.08 & 6.1 &0.25 & 6.0 &0.25 & 3.9 &0.16 & 13.5 &0.56 & 7.3 & 0.30 \\
\multicolumn{12}{c}{\textit{deprojected}}\\
2.4 & 0.10 & 7.8 & 0.32 & 7.6 & 0.32 & 5.0 & 0.21 & 17.1 & 0.71 & 9.3 & 0.39\\
\hline 
\\
& & \multicolumn{2}{c}{min} &\multicolumn{2}{c}{max} & \multicolumn{2}{c}{mean} & \multicolumn{2}{c}{median} & \multicolumn{2}{c}{std} \\
\cmidrule(lr){3-4}
\cmidrule(lr){5-6}
\cmidrule(lr){7-8}
\cmidrule(lr){9-10}
\cmidrule(lr){11-12}
& & [$\arcsec$] & [pc] & [$\arcsec$] & [pc] & [$\arcsec$] & [pc] & [$\arcsec$] & [pc] & [$\arcsec$] & [pc] \\

\multicolumn{2}{c}{\textit{projected}} & 1.9 & 0.08 & 13.5 & 0.56 & 6.5 & 0.27 & 6.1 & 0.25 & 3.9 & 0.16  \\
\multicolumn{2}{c}{\textit{deprojected}} &2.4 & 0.10 & 17.1 & 0.71 & 8.2 & 0.34 & 7.7 & 0.32 & 5.0 & 0.20  \\
\hline\hline\\

\end{tabular}
\newline
\flushleft
 {\small{\textbf{Notes}: Upper part: projected and deprojected (using the factor 4/$\pi$) angular and linear separations between cores along the arch. Lower part: minimum, maximum, mean, median, and standard deviation of the values reported in the upper part of the table.}}

\label{tab:separation}
\end{table*}
\subsection{Mass estimates and virial parameter}
\label{sect:mass}
To estimate the core masses we use the equation:
\begin{equation}
    M = \frac{100 \,F_{\rm{INT}}\, D^2}{B_{\nu}(T)\kappa_{\nu}},
\end{equation}
where $F_{\rm{INT}}$ is the integrated flux density derived from CuTEx at 219 GHz for cores from \#2 to \#6 and at 150 GHz for cores \#7 to \#9, $D$ is the distance of the source, $B_\nu(T)$ is the Planck function either at 219\,GHz or 150 GHz for the core dust temperature $T$, the factor 100 is the usually assumed gas-to-dust ratio, and $\kappa_{\nu}$ is the dust 
mass absorption coefficient, assumed to be 0.9\,cm$^2$\,g$^{-1}$ at 219\,GHz as in \cite{ALMAGAL3}, and scaled with a $\beta$ of 1.7 to derive the dust opacity coefficient at 150\,GHz, following the relation:
\begin{equation*}
    \kappa_{150} = \kappa_{219}\,\Biggl( \frac{150\,\rm{GHz}}{219\,\rm{GHz}}\Biggr) ^\beta\,.
\end{equation*}
 \indent For core \#4 we used as dust temperature the temperature estimated from the fit of CH$_3$CCH, assuming that gas and dust are coupled. For the other cores, under the same assumption, we use the ranges of temperature defined in section 4.2. We derived the mass ranges for those temperature ranges, and we listed them in Table \ref{tab:massvirial}. For these calculation we have used the usually adopted gas-to-dust ratio of 100, but as highlighted by the work of \cite{giannetti2017} this ratio reaches higher values in the outer Galaxy, even though the data show large scatter. In particular, at the galactocentric distance of \source \,    ($\sim10$\,kpc) the gas-to-dust ratio from the relation found in \cite{giannetti2017} would be of $\sim200$. Thus, these masses could be underestimated up to a factor two from the uncertainties on this parameter.\\ \indent In 
Table \ref{tab:massvirial} we also list an average FWHM from fits to all the molecular lines detected on the cores (listed in Table \ref{tab:moleculardetection}), obtained with a Gaussian fit of the mean spectrum over the size of each core. The fit to derive the FWHM for the three deuterated species reported in Table \ref{tab:moleculardetection} does not take into account the hyperfine splitting of the lines, therefore these are upper limits of the value of FWHM. We did not considered the hyperfine splitting because only in a few cases more transitions other than the main were detected in the spectrum, thus making the results of the fit including the hyperfine splitting not well constrained.
With these information we can derive the virial parameter of the cores, $\alpha_{\rm{vir}}$, defined as \citep{bertoldi1992}:
\begin{equation}
\alpha_{\rm{vir}} = \frac{2\,E_{\rm{kin}}}{E_{\rm{grav}}} = \frac{5\,\sigma^2\,r}{G\,M}\,,
\end{equation}
where $E_{\rm{kin}}$ and $E_{\rm{grav}}$ are the kinetic and gravitational potential energy of the cores, while $\sigma$ is the velocity dispersion, related to the FWHM by $\sigma=\mathrm{FWHM}/2.35$. The values are reported in Table 
\ref{tab:massvirial} and vary between 0.3 and 2.4, for all cores except core \#7 which has a virial parameter larger than 3. Despite the large ranges due to uncertainties in the temperature and therefore mass of the cores, all the cores except core \#7 are consistent with bound objects ($\alpha_{\rm{vir}}<2$, e.g. \citealt{kauffmann2013} ), not only cores \#3 and \#4 that already have a central 
condensation at $\sim1400\,$au. Moreover, as stated above, the FWHM and therefore $\sigma$ for most of the cores are upper limits of the real value, since we did not take into account the hyperfine splitting, while the masses could be lower limits if the real gas-to-dust ratio in the region is above 100. Therefore the values of $\alpha_{\rm{vir}}$ reported here are also upper limits. 
\section{Triggered star-formation?}
\begin{figure}
    \centering
    \includegraphics[width=1\linewidth, trim={1.3cm, 0, 1.cm ,0}, clip=True]{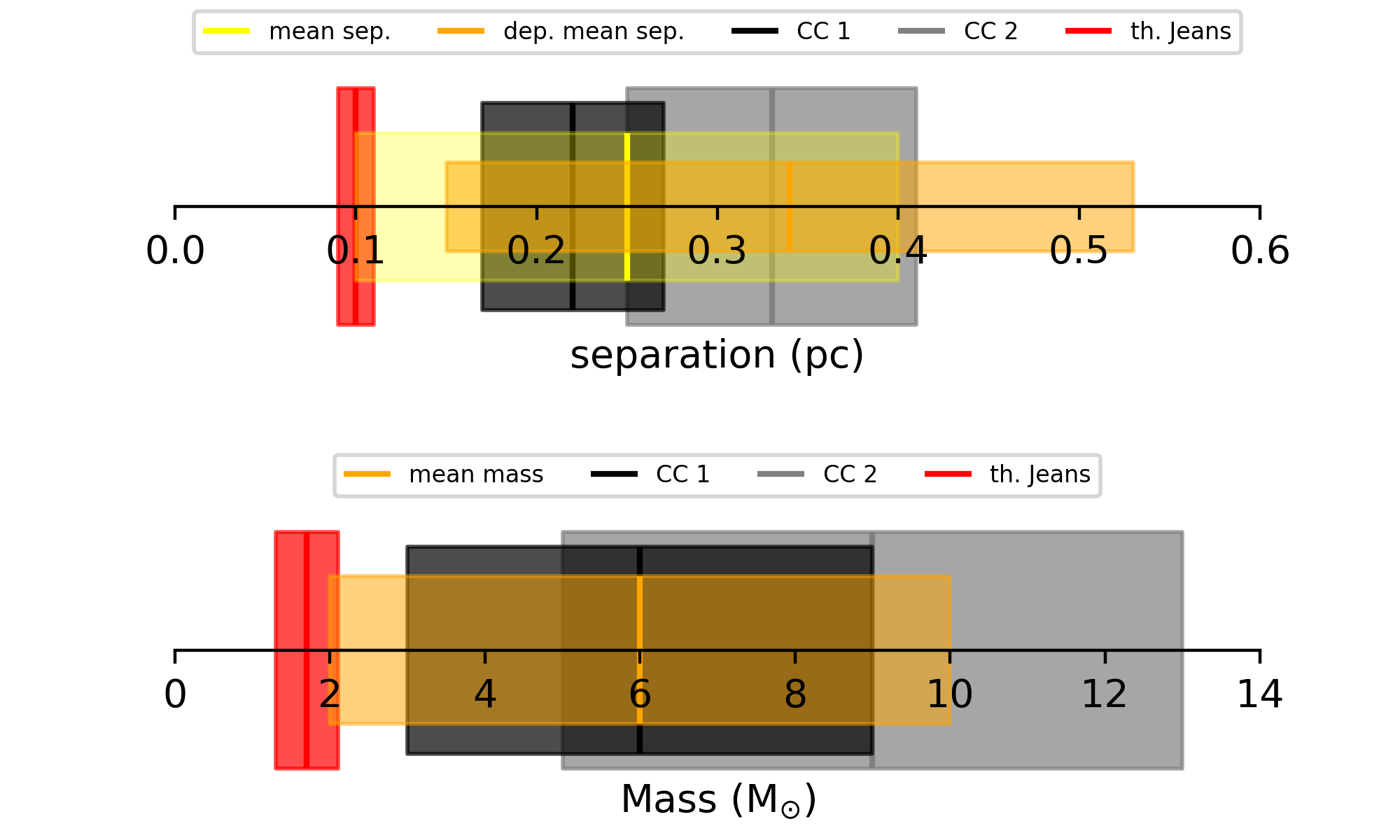}
    \caption{Upper panel: mean observed separation in yellow, and mean deprojected separation in orange. The black, gray, and red box represents the values derived from the collect and collapse model with $n_0=5.1\times10^{4}\,$cm$^{-3}$ (CC\,1), with  $n_0=2\times10^{4}\,$cm$^{-3}$ (CC\,2), and from the thermal Jeans fragmentation model, respectively. Lower panel: mean mass of the fragments in orange. The black, gray, and red boxes represents the values derived from the same models of the upper panel.}
    \label{fig:results}
\end{figure}

The presence of an \ion{H}{ii} region in our FOV and the location of the cores on an arched-shaped region centered at the \ion{H}{ii} region does not ensure that these cores are the result of triggered star formation, as found by \cite{Zhang2024ATOMS}. In the case of source \source, we also see from Fig. \ref{allbands} that there is a gap between the 23\,cm emission and the dust and gas arched structure, with no evidence of a photodissociation region (PDR), such as a bright rim in the images of IRAC band 1 at 3.6\,$\mu$m   (see Fig. D.1) due to the emission of polycyclic aromatic hydrocarbons (PAHs), while the data from IRAC band 3 and 4 are not available and other typical molecular tracers of PDRs \citep[see e.g.][]{kim2020} are not present in our data. One possibility is that the \ion{H}{ii} region is more extended than the region emitting at 23\,cm \citep{kim&koo2001,kurtz2002}, but the outer part of the \ion{H}{ii} is not efficiently emitting, due to possible lower values of the electron density or electron temperature. Another possibility is that the region is shaped not only by the \ion{H}{ii} region, but also by the presence of stellar winds \citep[see e.g.][]{kim2025}.\\
\indent In both cases, to test if the formation of these cores is triggered we compare the scale of fragmentation  and the typical mass of fragments along the arch with the value expected in the case of a shell fragmentation as prescribed by the collect and collapse process. The formulae to calculate these parameters have been derived by  \cite{whitworth1994shell}:
\begin{equation}
\lambda_{\rm{shell}} \sim 0.83\, \Big( \frac{c_s}{0.2 \rm{km/s}} \Big)^{18/11}\, \Big( \frac{N_{\rm{Ly}}}{10^{49} \rm{s^{-1}}} \Big)^{-1/11}\, \Big( \frac{n_0}{10^{3} \rm{cm^{-3}}} \Big)^{-5/11}\,\mathrm{pc}\,,
\end{equation}
\begin{equation}
M_{\rm{shell}} \sim 23\, \Big( \frac{c_s}{0.2 \rm{km/s}} \Big)^{40/11}\, \Big( \frac{N_{\rm{Ly}}}{10^{49} \rm{s^{-1}}} \Big)^{-1/11}\, \Big( \frac{n_0}{10^{3} \rm{cm^{-3}}} \Big)^{-5/11}\,\mathrm{M_{\odot}}\,,
\end{equation}
\begin{equation}
R_{\rm{shell}} \sim 5.8\, \Big( \frac{c_s}{0.2 \rm{km/s}} \Big)^{4/11}\, \Big( \frac{N_{\rm{Ly}}}{10^{49} \rm{s^{-1}}} \Big)^{1/11}\, \Big( \frac{n_0}{10^{3} \rm{cm^{-3}}} \Big)^{-6/11}\,\mathrm{pc}\,,
\end{equation}
\begin{equation}
t_{\rm{shell}} \sim 1.56\, \Big( \frac{c_s}{0.2 \rm{km/s}} \Big)^{7/11}\, \Big( \frac{N_{\rm{Ly}}}{10^{49} \rm{s^{-1}}} \Big)^{-1/11}\, \Big( \frac{n_0}{10^{3} \rm{cm^{-3}}} \Big)^{-5/11}\, \mathrm{yr}\,,
\end{equation}
where $\lambda_{\rm{shell}}$, $M_{\rm{shell}}$, $R_{\rm{shell}}$, and $t_{\rm{shell}}$ are the characteristic scale of fragmentation, the characteristic mass of the fragments, and the radius of the shell at the time of fragmentation $t_{\rm{shell}}$, respectively.\\ \indent The only quantities needed to derive these parameters are the rate of ionizing photons $N_{\rm{Ly}}\sim 6.2\times10^{46}$phot.\,s$^{-1}$, as derived in Sect. 3.2, the  isothermal sound speed for the shell $c_s$, and the initial density of the material swept by the expanding \ion{H}{ii} region, $n_0$, assuming that the mean density at clump scale observed is a good estimate of its initial value before the expansion of the \ion{H}{ii} region. We used the values derived from the Herschel infrared Galactic Plane Survey (Hi-Gal) \citep{elia2021}, i.e. from the fit of the infrared SED at clump scale ($\sim$1\,pc), for the temperature of the clump, its mass, and radius to derive the best estimate of $c_s$ and $n_0$. Those are 0.2 km\,s$^{-1}$ and 5.1$\times10^4$cm$^{-3}$, respectively. Using these values, and assuming a 10\% error for all the three parameters in Eq.s (5)-(8), we find  $\lambda_{\rm{shell}}\sim 0.22\pm0.05$\,pc, $M_{\rm{shell}}\sim6\pm3$\,M$_{\odot}$, $R_{\rm{shell}}\sim0.43\pm0.04$\,pc, and $t_{\rm{shell}}\sim0.42\pm0.05$\,Myr.\\ \indent We list the separation of all the bound cores along the arch (i.e. all except core \#7) in Table \ref{tab:separation}. The median values in \source\, is $0.25\pm0.15$\,pc. However, this is a lower limit of the separation, assuming that all the cores are perfectly aligned on the plane of the sky. For a couple of randomly oriented sources in 3d, the average deprojection factor is 4/$\pi$ (see \citealt{schisano2025}). The mean deprojected value is  $0.34\pm0.19$. We consider this value an upper limit for the mean separation, since from the disposition of the cores on a circular arch it is likely that the 3d structure is not randomly oriented, and the 3d separations do not have large contribution in the direction perpendicular to the plane of the sky. Both the deprojected and the observed mean separation of the cores are compatible within the error to that expected from the collect and collapse scenario, with the deprojected value being a factor $\sim 1.5$ larger than the estimate for the model. The mean mass of the cores considering both the upper limits and the lower limits reported in Table \ref{tab:massvirial} is $6\pm4$\,M$_{\odot}$. Therefore they are also compatible with the characteristic mass $M_{\rm{shell}}\sim6$\,M$_{\odot}$, while  $R_{\rm{shell}}\sim0.4$\,pc is smaller than the radius of the arc of 0.75\,pc, but within a factor 2. \\ \indent Assuming the same values of $c_s$ and $N_{\rm{Ly}}$, a radius of $\sim0.75\,$pc from Eq. (7) corresponds to a value of the initial volume density before the expansion of the \ion{H}{ii} region $n_0\sim2\times10^4\,$cm$^{-3}$, only a factor $\sim2.5$ smaller than the value evaluated by \cite{elia2021} for the current mean density of the clump. Using this new value of $n_0$ the estimates from Eq.s (5), (6), and (8) would be $\lambda_{\rm{shell}}\sim 0.34\pm0.08$\,pc, $M_{\rm{shell}}\sim9\pm4$\,M$_{\odot}$, and $t_{\rm{shell}}\sim0.67\pm0.08$\,Myr. The fragmentation scale and the typical mass of the fragments would still be compatible with the median values observed within their errors, differing of a factor $\sim1.1$ and $\sim1.5$ from the observed values respectively.\\ \indent 
We compare the mean separation and mass of the fragments also with the thermal Jeans length and the thermal Jeans mass, using the properties at clump level. These have been derived by \cite{schisano2025} to be 0.10\,pc and 1.7\,M$_{\odot}$, with an error of 0.01\,pc and 0.4\,M$_{\odot}$ respectively assuming a 10\% error on the two parameters involved in their calculation, i.e. the volume density and the temperature of the gas, as seen in Eq.s (B.2) and (B.3) of \cite{schisano2025}.\\ \indent The thermal Jeans length is a factor 2.5 lower than the mean observed separation, and a factor 3.4 lower than the mean deprojected separation and not compatible within the errors. The thermal Jeans mass is a factor $\sim3.5$ lower than the mean of the masses observed. The thermal Jeans mass is marginally compatible within the error with the mean mass from observations (upper limit from the model: $1.7+0.4=2.1\,\mathrm{M}_{\odot}$; lower limit of the mean of the mass estimated $6-4=2\,\mathrm{M}_{\odot}$). Nevertheless, as discussed in Sect. \ref{sect:mass} the dust-to-gas ratio in this region could be higher than 100, leading to higher estimates of the masses observed. \\ \indent
Figure \ref{fig:results} summarizes the comparison between observations and models, showing that the observed separations and masses for the cores in \source\, are in agreement with predictions from the collect and collapse models.

\section{Conclusions}
To study the nature of the disappearing cores, i.e. cores identified as compact sources in the ALMAGAL \textsc{7m+tm2} continuum at $\sim5000$\,au (7600\,au for \source) with no counterpart in the ALMAGAL \textsc{7m+tm2+tm1} continuum at $\sim1000$\,au (1400\,au for \source), we collected archival continuum centimeter and infrared data and new molecular line observations of the source \source, which is associated with a candidate \ion{H}{ii} region from the catalog of \cite{anderson2014}. 
Here we summarize the main results of our work:
\begin{itemize}
\item We built the radial profiles of the emission of the cores in the ALMAGAL FOV and confirmed that the profiles of the disappearing cores are shallower than those of cores \#3 and \#4, except core \#6. This core  has a profile similar to those of core \#3 and \#4, which are detected also in the higher resolution ALMAGAL images. The flux density radial profiles are compatible with the flux density expected for Bonnor-Ebert density profiles, with a larger inner flat region for the disappearing cores.
    \item From the analysis of the radial profiles and of the emission of \dco (2$-$1), \ntwod (2$-$1), and SiO $(5-4)$ we confirm that cores \#2, \#5, and \#6 (i.e. 3 out of 4 disappearing cores) are young prestellar objects. The other disappearing core, core \#1, does not have a compact morphology when observed in ALMA Band 4, and it likely corresponds to material dispersed by the expansion of the \ion{H}{ii} region or the relic of the dust envelope around the ionizing star.
    \item From the $\sim$23\,cm radio continuum emission we inferred the spectral index between 0.8\,GHz and 1.6\,GHz, which is consistent with the flat index of optically thin free-free emission ($\alpha\sim-0.1$ to $-0.4$). We therefore confirmed the presence of a \ion{H}{ii} region in this source. We derived a rate of ionizing photons $N_{\rm{Ly}}\sim6.2\times10^{46}$\,photons\,s$^{-1}$, which corresponds to a B0.5 star.
    \item The continuum emission in ALMA Bands 4 and 6 revealed a distribution of cores along an arched shaped region of radius $\sim0.75$\,pc centered close to the peak of the 23\,cm radio continuum emission.
    \item From the chemical characterization of the cores we derived a evolution sequence among the cores and assigned a range of gas and dust temperatures. Only towards the most evolved core \#4 we were able to derive a precise measurement of the temperature (39$\pm$5\,K), thanks to the detection of CH$_3$CCH. 
    \item The virial parameter derived for all the cores in the arched structure is consistent with bound objects, with the exception of core \#7.
    \item We compared the separations of the cores along the arch structure and their typical masses with the characteristic separation and mass derived in the collect and collapse scenario and with the thermal Jeans length and mass. The best agreement is found with the characteristic scales in the case of triggered star formation. 
\end{itemize}
This study confirmed that disappearing cores are good prestellar cores candidates. The presence of several of these cores in the ALMAGAL sample will allow us to increase the statistics and study how these early stages are affected by the feedback of already formed massive stars.

\begin{acknowledgements}
The authors thank K. Furuya for the plot that shows the result for \dco of the D/H drop after the temperature increase in the model presented in Fontani et al. 2014. The authors thanks the referee for their insightful comments.
C.M. and the INAF-IAPS team acknowledge funding from the European
Research Council (ERC) under the European Union's Horizon 2020 program
through the ECOGAL Synergy grant (ID 855130). CM acknowledges funding from INAF Mini Grants RSN2 2024 "Zodyac" CUP C83C25000340005. RK acknowledges financial support via the Heisenberg Research Grant funded by the Deutsche Forschungsgemeinschaft (DFG, German Research Foundation) under grant no.~KU 2849/9, project no.~445783058. Part of this research was carried out at the Jet Propulsion Laboratory, California Institute of Technology, under a contract with the National Aeronautics and Space Administration (80NM0018D0004). D.C.L. acknowledges financial support from the National Aeronautics and Space Administration (NASA) Astrophysics Data Analysis Program (ADAP). L.B. gratefully acknowledges support by the ANID BASAL project FB210003. A.S-M. acknowledges support from the PID2023-146675NB grant funded by MCIN/AEI/10.13039/501100011033, and by the programme Unidad de Excelencia Mar\'{\i}a de Maeztu CEX2020-001058-M, as well as support from the RyC2021-032892-I grant funded by MCIN/AEI/10.13039/501100011033 and by the European Union `Next GenerationEU'/PRTR. RSK acknowledges financial support from the ERC via Synergy Grant ``ECOGAL'' (project ID 855130) and from the German Excellence Strategy via the Heidelberg Cluster ``STRUCTURES'' (EXC 2181 - 390900948). In addition RSK is grateful for funding from the German BMWE in project ``MAINN'' (funding ID 50OO2206), and from DFG and ANR for project ``STARCLUSTERS'' (funding ID KL 1358/22-1).
This paper makes use of the following ALMA data:
project code 2019.1.00195.L and 2023.1.01386.S. ALMA is a partnership of ESO (representing
its member states), NSF (USA) and NINS (Japan), together with NRC (Canada), MOST and ASIAA (Taiwan), and KASI (Republic of Korea), in cooperation with the Republic of Chile. 
The Joint ALMA Observatory is operated by ESO,
AUI/NRAO and NAOJ.
\end{acknowledgements}

\bibliographystyle{aa}

\appendix
\onecolumn
\section{Spectra averaged on the cores}
\begin{figure*}
    \centering
    \includegraphics[width=12.0cm, trim={3.5cm, 0cm, 0cm, 0}, clip=True]{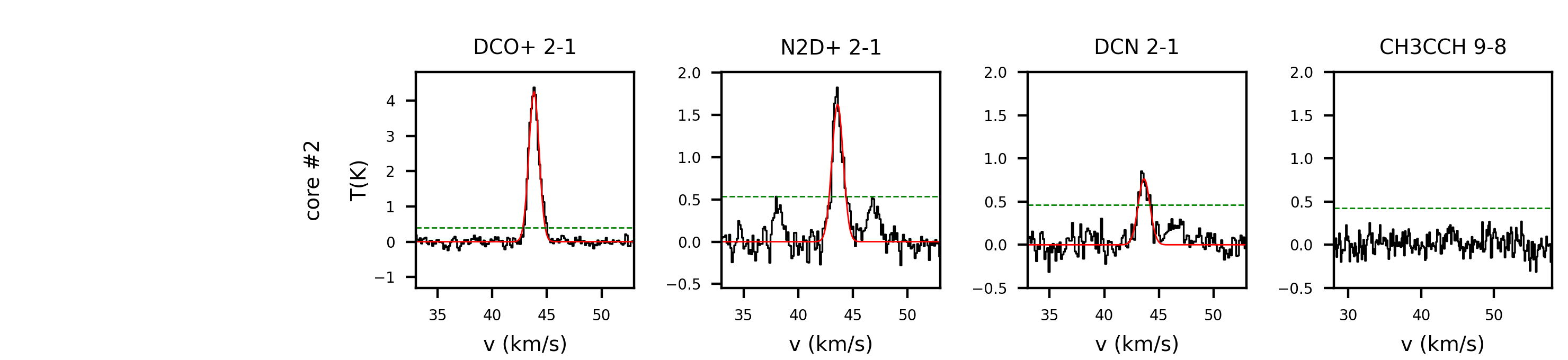}\\
    \includegraphics[width=12.0cm, trim={3.5cm, 0cm, 0cm, 0}, clip=True]{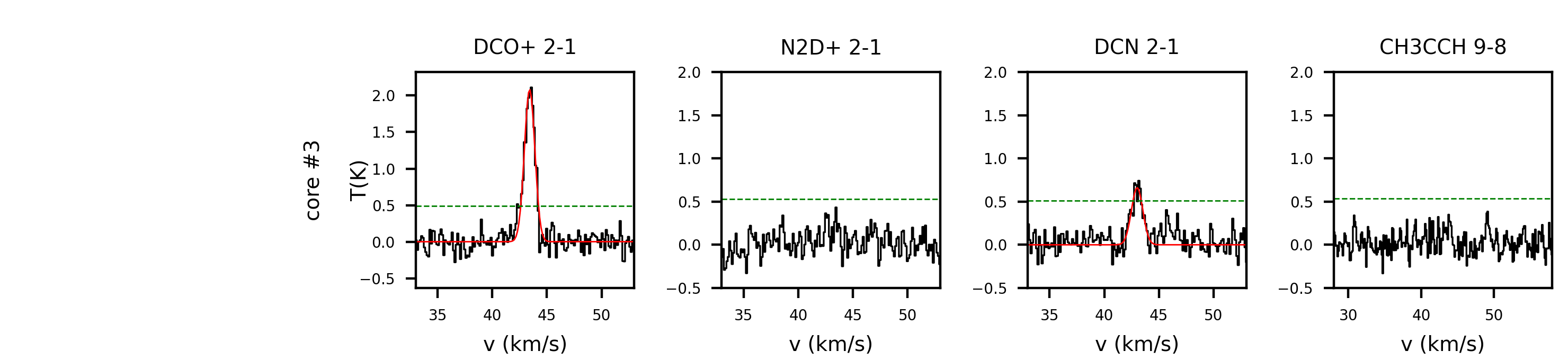}\\
    \includegraphics[width=12.0cm, trim={3.5cm, 0cm, 0cm, 0}, clip=True]{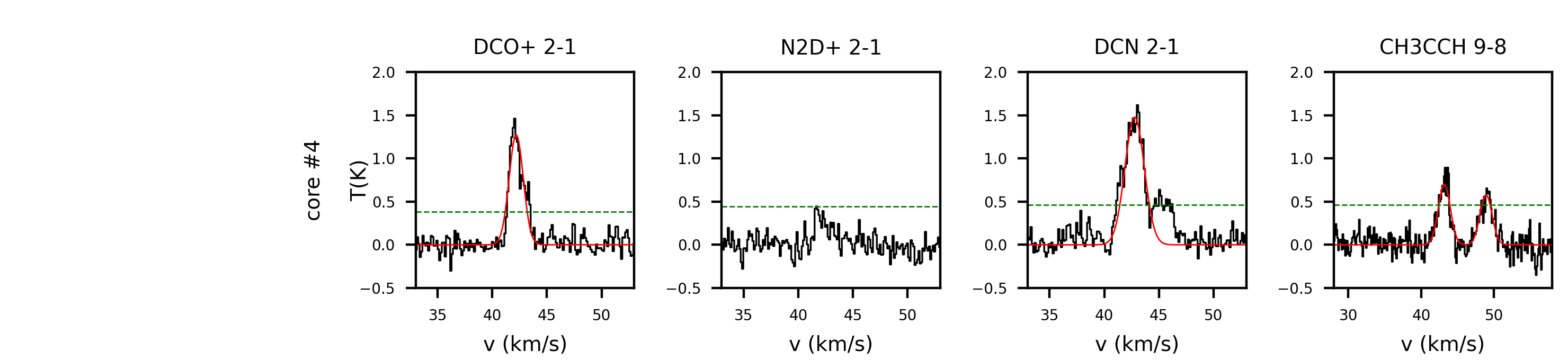}\\
    \includegraphics[width=12.0cm, trim={3.5cm, 0cm, 0cm, 0}, clip=True]{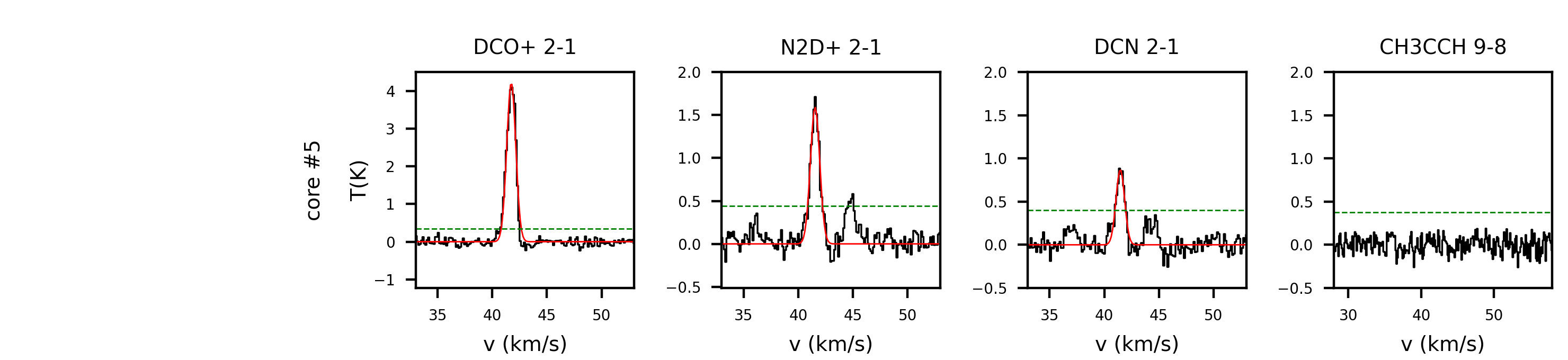}\\
    \includegraphics[width=12.0cm, trim={3.5cm, 0cm, 0cm, 0}, clip=True]{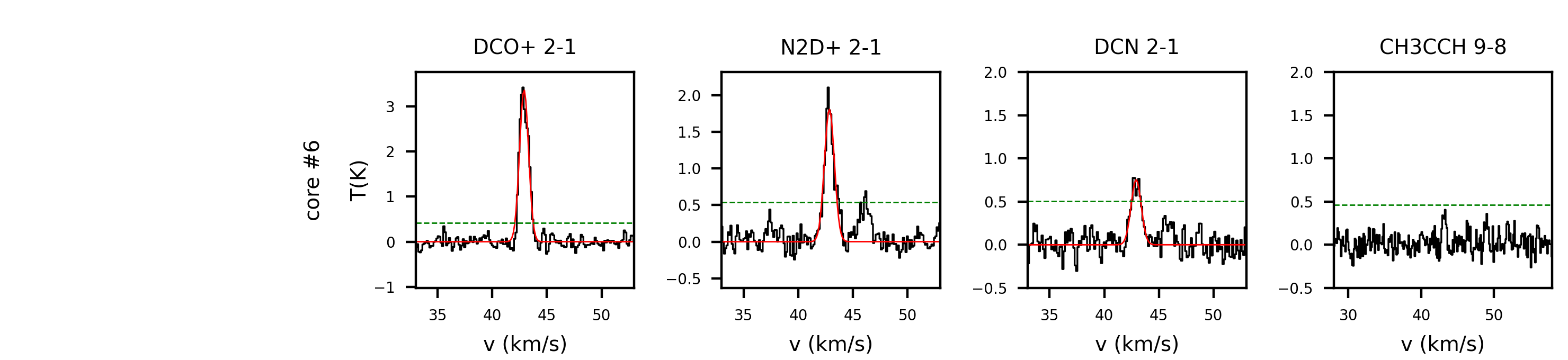}
    \includegraphics[width=12.0cm, trim={3.5cm, 0cm, 0cm, 0}, clip=True]{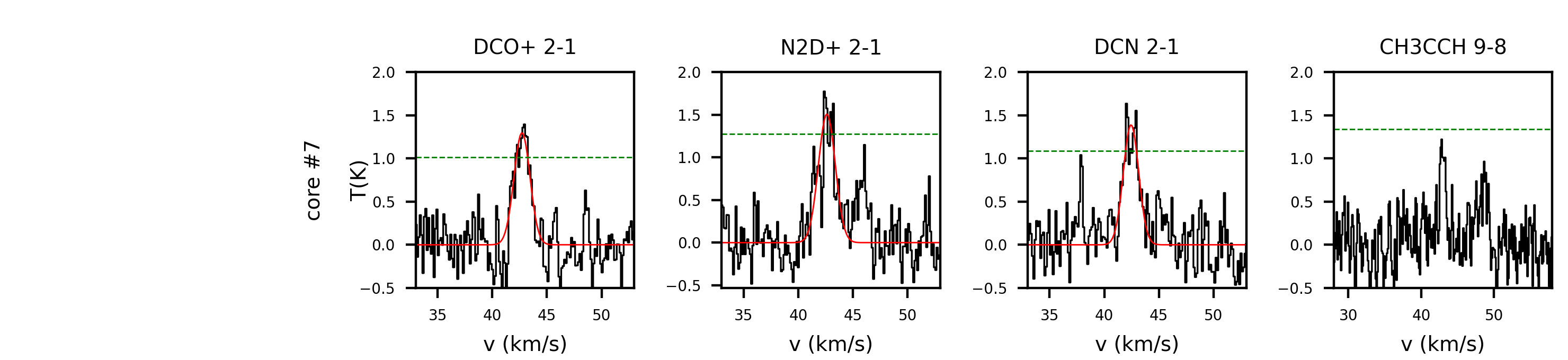}\\

    \caption{Spectra averaged over the cores (from top to bottom: core \#2 to \#7). The dashed horizontal line is 5 time the rms of the spectra.}
    \label{fig:detection}
\end{figure*}

\begin{figure*}
    \centering
    \includegraphics[width=12.0cm, trim={3.5cm, 0cm, 0cm, 0}, clip=True]{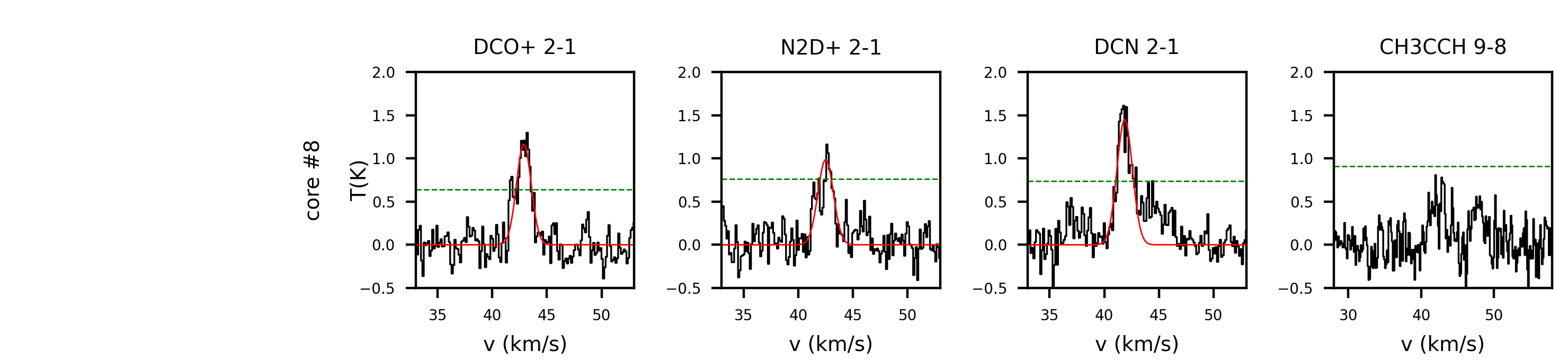}\\
    \includegraphics[width=12.0cm, trim={3.5cm, 0cm, 0cm, 0}, clip=True]{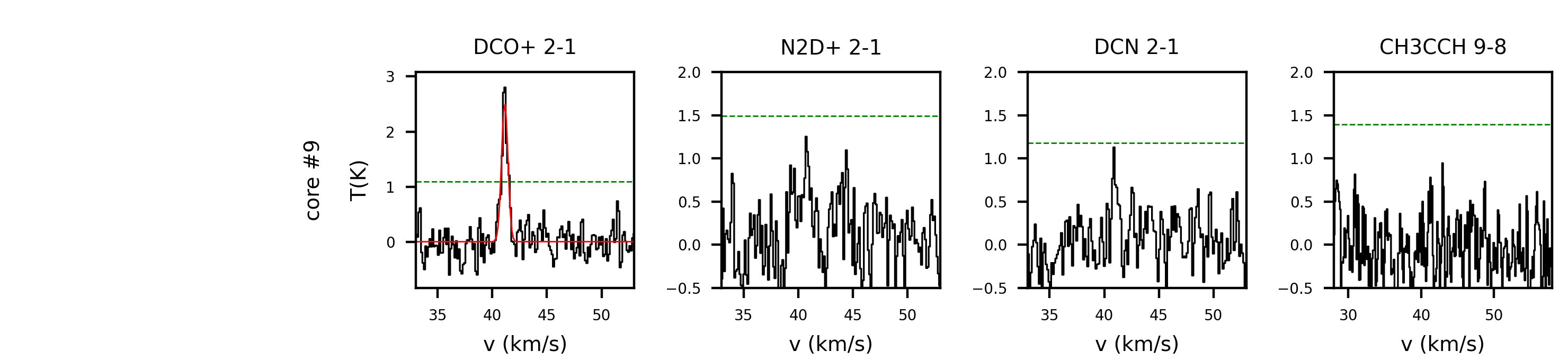}\\

    \caption{Spectra averaged over the cores (from top to bottom: core \#8 to \#9). The dashed horizontal line is 5 time the rms of the spectra.}
    \label{fig:detection2}
\end{figure*}

For each cores detected in the ALMA data towards \source, we extracted the average spectra for \ntwod (2--1), \dco (2--1), DCN (2--1), and CH$_3$CCH (9--8). These are shown in Fig. \ref{fig:detection} and \ref{fig:detection2}.

\section{D/H ratio drop timescale for \dco}
\begin{figure}
    \centering
    \includegraphics[width=0.99\linewidth]{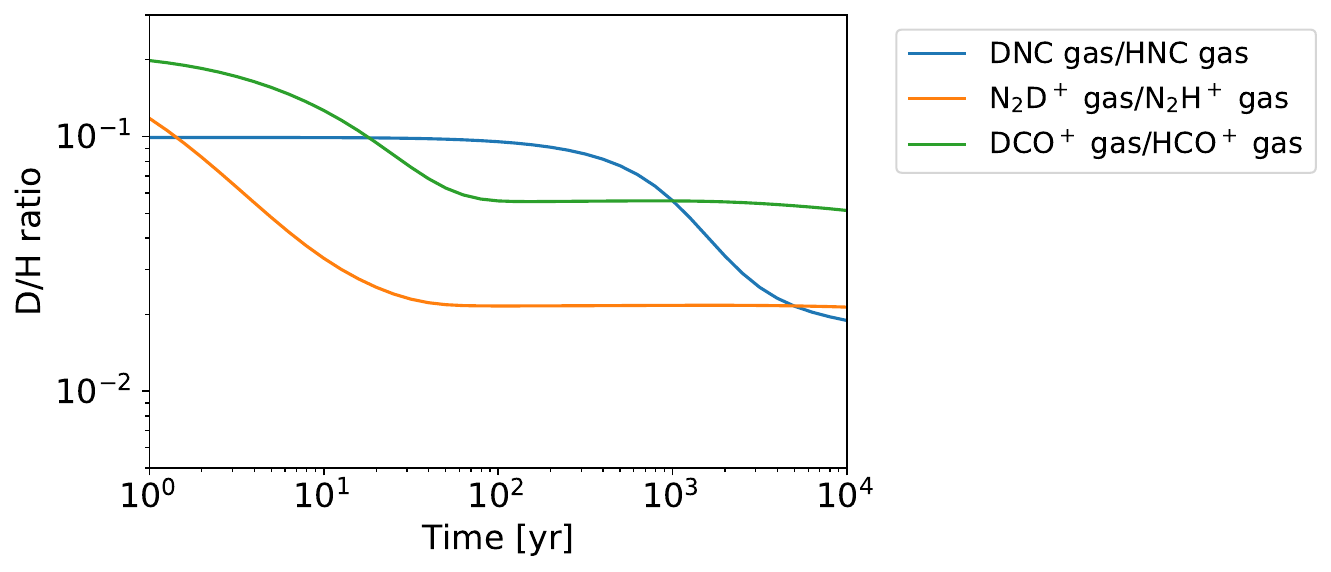}
    \caption{D/H ratio as a function of time after the temperature rises from 15\,K to 40\,K.}
    \label{dcodrop}
\end{figure}

In Fig. \ref{dcodrop} we show the results of the model by \cite{fontani2014timescale} in the case of a density of 10$^{5}$\,cm$^{-3}$ (private communication with F. Fontani and K. Furuya). 

\section{SiO moment maps}
\begin{figure*}
    \centering
    \includegraphics[width=18cm, trim={0cm, 0cm, 0cm, 0}, clip=True]{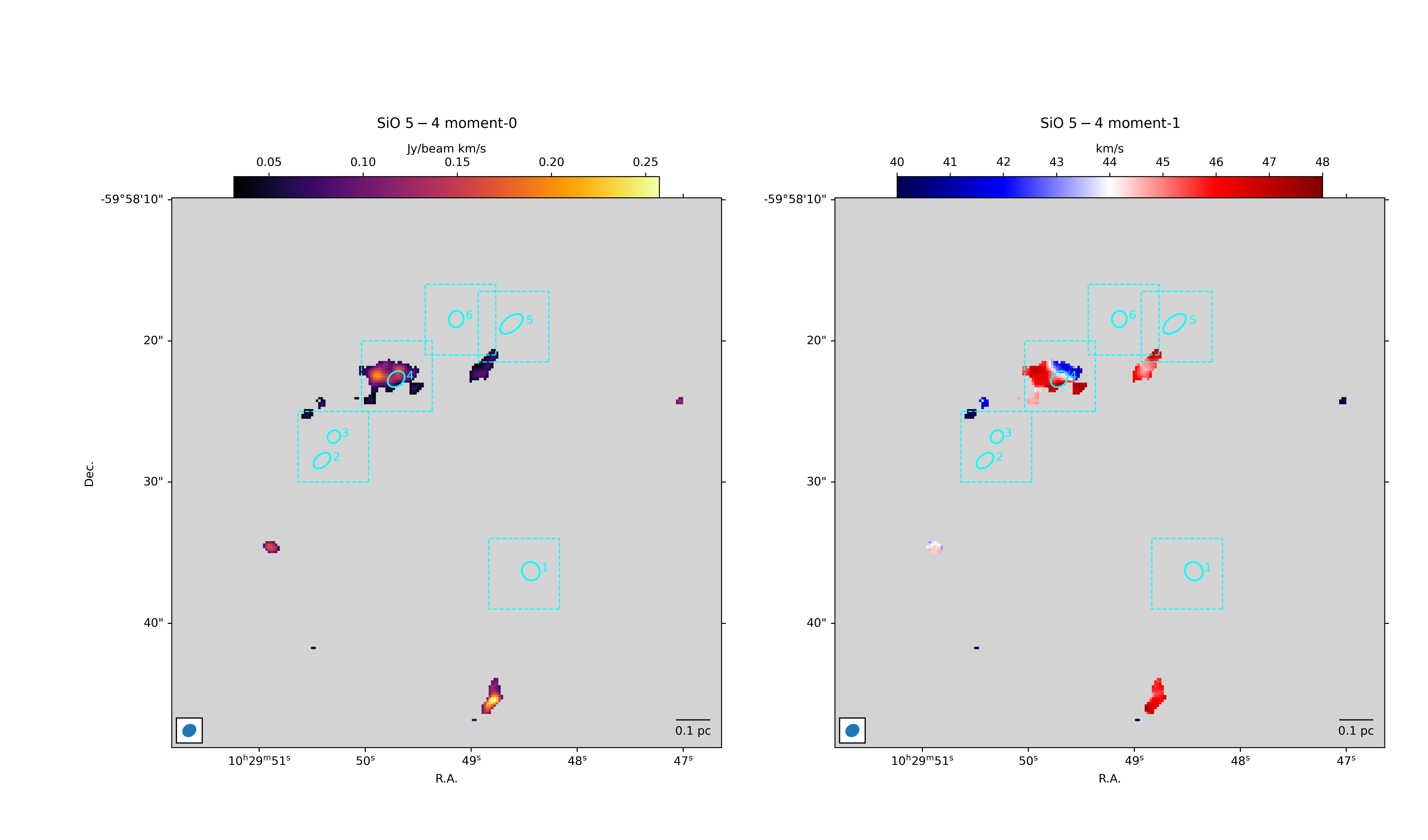}
    \caption{Moment-0 map (left) and moment-1 map (right) of the SiO $(5-4)$ transition, from ALMAGAL data. The cyan ellipses represent the cores identified by CuTEx in the 219\,GHz \textsc{7m+tm2} continuum data. }
    \label{fig:SiO}
\end{figure*}

We created the moment-0 and moment-1 maps of SiO (5-4) in the ALMAGAL dataset using the same methodology described in Sect. 3.3. These are shown in Fig. \ref{fig:SiO}
\section{IRAC emission}
\begin{figure*}
    \centering
    \includegraphics[width=18cm, trim={0cm, 0cm, 10.5cm, 0}, clip=True]{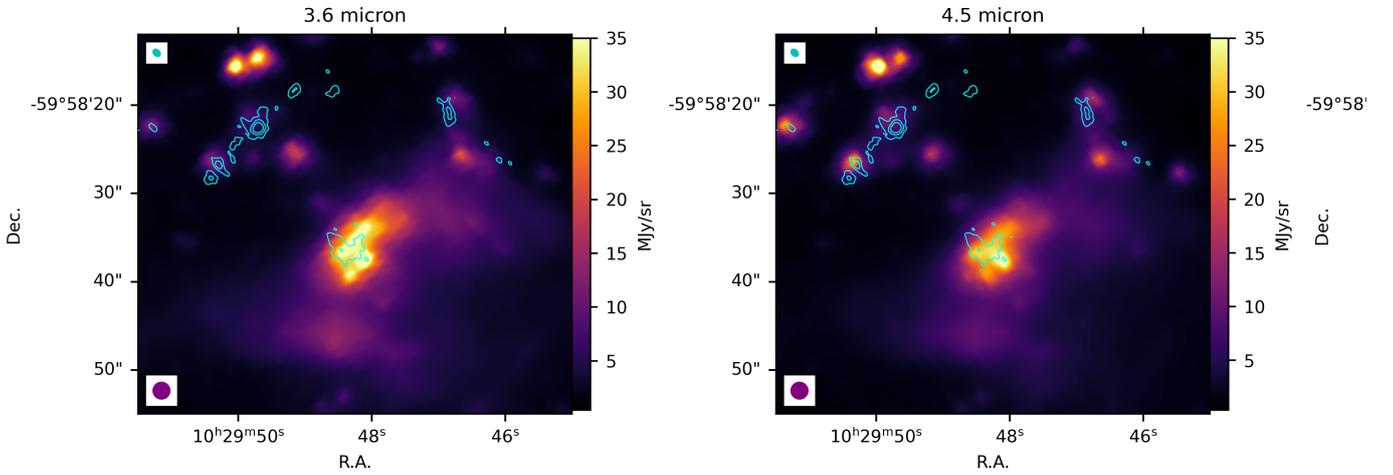}
    \caption{Emission at the coordinates of source \source\, in IRAC band 1 and band 2. The cyan contours are the contour of the continuum emission in ALMA Band 4 (4, 10, and 20 times the rms). The IRAC beam is shown in the bottom-left corner in purple, while the ALMA Band 4 beam is shown in cyan in the upper-left corner. }
    \label{fig:placeholder}
\end{figure*}

\end{document}